\documentclass[twocolumn,longauth]{aa} % for a paper on 1 column  
\usepackage{aalongtable}
\usepackage{subfigure}  %% permette di impostare piu' figure 

\usepackage{natbib}
\usepackage{txfonts}
\usepackage{lineno}
\usepackage[english]{babel}
\usepackage{graphicx}
\usepackage{color}

\newcommand{\gsim}{\;\lower.6ex\hbox{$\sim$}\kern-7.75pt\raise.65ex\hbox{$>$}\;}
\newcommand{\lsim}{\;\lower.6ex\hbox{$\sim$}\kern-7.75pt\raise.65ex\hbox{$<$}\;}
\def\lHa{H${\alpha}\lambda$6563$\AA$}
\def\lHb{H${\beta}\lambda$4861$\AA$}
\def\Ha{H${\alpha}$}
\def\Hb{H${\beta}$}

\def\NII{[N\,\textsc{ii}]}
\def\SII{[S\,\textsc{ii}]}
\def\lNII{[N\,\textsc{ii}]$\lambda$6583$\AA$}
\def\lSII{[S\,\textsc{ii}]$\lambda\lambda$6717,6731$\AA$}

\def\lOI{[O\,\textsc{i}]$\lambda$6300$\AA$}

\def\lOIIIb{[O\,\textsc{iii}]$\lambda$5007$\AA$}
\def\OIIIb{[O\,\textsc{iii}]}
\def\OII{[O\,\textsc{ii}]}
\def\NeIII{[Ne\,\textsc{iii}]}
\def\lOII{[O\,\textsc{ii}]$\lambda$3727$\AA$}

\def\reda{[N\,\textsc{ii}]/H${\alpha}$}
\def\redb{[S\,\textsc{ii}]/H${\alpha}$}
\def\blue{[O\,\textsc{ii}]/H${\beta}$}  
\begin{document}
%\linenumbers*                                               
%
   \title{The \OIIIb\ emission line luminosity function of optically 
   selected type--2 AGN from zCOSMOS \footnote{based on data obtained with the European 
   Southern Observatory Very Large Telescope, Paranal, Chile, program 175.A-0839}}

\author{
A. Bongiorno \inst{1,2} 
\and M. Mignoli \inst{3}  
\and G. Zamorani \inst{3} 
\and F. Lamareille \inst{4}
\and G. Lanzuisi \inst{5,6} 
\and T. Miyaji\inst{7,8} 
\and M. Bolzonella\inst{3} 
\and C. M. Carollo \inst{9}
\and T. Contini \inst{4}
\and J. P. Kneib \inst{10}
\and O. Le F\`evre  \inst{10}
\and S. J. Lilly  \inst{9} 
\and V. Mainieri\inst{11}
\and A. Renzini\inst{12} 
\and M. Scodeggio \inst{13} 
\and S. Bardelli \inst{3} 
\and M. Brusa \inst{1} 
\and K. Caputi \inst{9} 
\and F. Civano \inst{5} 
\and G. Coppa \inst{3,14} 
\and O. Cucciati \inst{10} 
\and S. de la Torre \inst{10,14,16} 
\and L. de Ravel  \inst{10} 
\and P. Franzetti \inst{13}
\and B. Garilli \inst{13}
\and C. Halliday\inst{16}
\and G. Hasinger\inst{1,17}
\and A. M. Koekemoer \inst{18} 
\and A. Iovino \inst{13} 
\and P. Kampczyk \inst{9} 
\and C. Knobel \inst{9}
\and K. Kova\v{c} \inst{9}
\and J. -F.~Le Borgne \inst{4}
\and V. Le Brun \inst{10} 
\and C. Maier \inst{9}
\and A. Merloni \inst{1,19}
\and P. Nair \inst{3} 
\and R. Pello \inst{4}
\and Y. Peng \inst{9}
\and E. Perez Montero \inst{4,20} 
\and E. Ricciardelli \inst{21} 
\and M. Salvato\inst{13,22} 
\and J. Silverman\inst{9}
\and M. Tanaka \inst{11}
\and L. Tasca\inst{10,13} 
\and L. Tresse  \inst{10}
\and D. Vergani \inst{3} 
\and E. Zucca \inst{3} 
\and U. Abbas \inst{10}
\and D.~Bottini \inst{13}
\and A.~Cappi \inst{3}
\and P.~Cassata \inst{10,23}
\and A.~Cimatti \inst{14}
\and L.~Guzzo \inst{16}
\and A.~Leauthaud \inst{10}
\and D.~Maccagni \inst{13}
\and C.~Marinoni \inst{24}
\and H.~J. McCracken \inst{25}
\and P.~Memeo \inst{13}
\and B.~Meneux \inst{1}
\and P.~Oesch \inst{9}
\and C.~Porciani \inst{9}
\and L.~Pozzetti \inst{3}
\and R.~Scaramella \inst{26}
}

\offprints{Angela Bongiorno, \email{angelab@mpe.mpg.de}}

\institute{Max-Planck-Institut f\"ur extraterrestrische Physik (MPE) 
- Giessenbachstra\ss e 1, D-85748, Garching bei M\"unchen, Germany
\and
University of Maryland, Baltimore County, 1000 Hilltop Circle, Baltimore, MD21250, USA
\and
 INAF-Osservatorio Astronomico di Bologna - Via Ranzani 1, I-40127, Bologna, Italy
\and
Laboratoire d'Astrophysique de Toulouse-Tarbes, Universit{\'e} de Toulouse, CNRS, 14 avenue Edouard Belin, F-31400 Toulouse, France
\and
Harvard-Smithsonian Center for Astrophysics, 60 Garden Street, Cambridge, MA 02138
\and
Dipartimento di Fisica, Universit\`a di Roma La Sapienza, P.le A. Moro 2, 00185 Roma, Italy
\and
Instituto de Astronom\'ia, Universidad Nacional Aut\'onoma de M\'exico, Ensenada, M\'exico, (mailing address: PO Box 439027, San Ysidro, CA, 92143-9027, USA)
\and
University of California, San Diego, Center for Astrophysics and Space Sciences, 9500 Gilman Drive, La Jolla, CA 92093-0424, USA
\and
Department of Physics, ETH Zurich, CH-8093 Zurich, Switzerland
\and
Laboratoire d'Astrophysique de Marseille, CNRS-Univerist{\'e} d'Aix-Marseille, 38 rue Frederic Joliot Curie, 13388 Marseille Cedex 13, France
\and
ESO, Karl-Schwarzschild-Strasse 2, D-85748 Garching bei M\"unchen, Germany
\and
INAF - Osservatorio Astronomico di Padova, Padova, Italy
\and
INAF - Istituto di Astrofisica Spaziale e Fisica Cosmica di Milano, via Bassini 15, I-20133, Milano, Italy
\and
Dipartimento di Astronomia, Universit{\`a} di Bologna, via Ranzani 1, I-40127, Bologna, Italy
\and
INAF Osservatorio Astronomico di Brera, Via Brera 28, I-20121 Milano, Italy
\and
Osservatorio Astrofisico di Arcetri, Largo Enrico Fermi 5, I-50125 Firenze, Italy
\and
Max-Planck-Institute f\"ur Plasmaphysik, Boltzmannstrasse 2, D-85748 Garching bei M\"unchen, Germany
\and
Space Telescope Science Institute, 3700 Martin Drive, Baltimore, MD 21218
\and
Excellence Cluster Universe, TUM, Boltzmannstr. 2, 85748, Garching bei M\"unchen, Germany
\and
Instituto de Astrofisica de Andalucia, CSIC, Apdo. 3004, 18080, Granada, Spain
\and
Dipartimento di Astronomia, Universit{\`a} di Padova, vicolo Osservatorio 3, I-35122 Padova, Italy
\and
California Institute of Technology, MC 105-24, 1200
 East California Boulevard, Pasadena, CA 91125, USA
\and
Department of Astronomy, University of Massachusetts, 710 North Pleasant Street, Amherst, MA 01003, USA
\and
Centre de Physique Theorique, Marseille, France
\and
Institut d'Astrophysique de Paris, Universit\'e Pierre \& Marie Curie, Paris, France
\and
INAF - Osservatorio Astronomico di Roma, via di Frascati 33, I-00040 Monteporzio Catone, Italy
}

\date{Received; accepted}

\abstract{}
{We present a catalog of 213 type--2 AGN selected from the zCOSMOS survey. 
The selected sample covers a wide redshift range (0.15$<z<$0.92) and 
is deeper than any other previous study, encompassing the luminosity range 
$10^{5.5}$L$_{\odot}<$L$_{\rm [OIII]}<$ 10$^{9.1}$ L$_{\odot}$.
We explore the intrinsic properties of these AGN and the relation to
their X-ray emission (derived from the XMM-COSMOS observations). We study their evolution 
by computing the \lOIIIb\ line luminosity function (LF) and we constrain the fraction 
of obscured AGN as a function of luminosity and redshift.}
{The sample was selected on the basis of  the optical emission line ratios, after applying a cut to 
the signal-to-noise ratio (S/N) of the relevant lines. We used the standard 
diagnostic diagrams (\OIIIb/\Hb\ versus \NII/\Ha\ and \OIIIb/\Hb\ versus \SII/\Ha) 
to isolate AGN in the redshift range 0.15$<z<$0.45 and the diagnostic diagram 
\OIIIb/\Hb\ versus \OII/\Hb\ to extend the selection to higher redshift (0.5$<z<$0.92).}
{Combining our sample with one drawn from SDSS, we found that the best description of the 
evolution of type--2 AGN is a luminosity-dependent density evolution model. 
Moreover, using the type--1 AGN LF we were able to constrain the fraction of type--2 AGN to the 
total (type--1 + type--2) AGN population. We found that the type--2 fraction decreases with luminosity, 
in agreement with the most recent results, and shows signs of a slight increase with redshift. 
However, the trend with luminosity is visible only after 
combining the SDSS+zCOSMOS samples. From the COSMOS data points alone, the type--2 fraction seems 
to be quite constant with luminosity.}
%According to our data, in the past $\sim$5 Gyr (0.15$<z<$0.92) at the faintest \OIIIb\ 
%luminosities, type--2 AGN constitute more than half of the AGN population.}
{}
% 5 {} token are mandatory


\titlerunning{The evolution of type--2 AGN from the zCOSMOS Survey}
\authorrunning{Bongiorno, A. et al.}

\maketitle
  
%
%________________________________________________________________
\section{Introduction}

According to the standard unified model \citep[e.g.;][]{Antonucci1993}, AGN can be broadly classified
into two categories depending on whether the central black hole and its associated continuum and
broad emission-line region are viewed directly (type--1 AGN) or are obscured by a dusty circumnuclear 
medium (type--2 AGN). 
Type--1 AGN are characterized by power-law continuum emission, broad permitted emission 
lines ($\gsim$1000 km s$^{-1}$) and are thus easily recognizable from their spectra. 
In contrast, type--2 AGN have narrow permitted and forbidden lines ($\lsim$1000 km s$^{-1}$) 
and their stellar continuum, often dominated by stellar emission, is similar to normal star-forming galaxies 
(SFGs).
The main difference between AGN and SFGs is the ionizing source responsible for their emission lines: 
non-thermal continuum 
from an accretion disc around a black hole for AGN or photoionization by hot massive stars 
for normal SFGs.

To identify type--2 AGN, we thus need to determine the ionizing source. 
\citet{Baldwin1981} demonstrated how this is possible by considering 
the intensity ratios of two pairs of relatively strong emission lines. 
In particular, they proposed a number of
diagnostic diagrams (hereafter BPT diagrams), which were further refined by \citet{Veilleux1987},
based on \lOIIIb, \lOI, \lNII, \lSII, \lHa\ and \lHb\ emission lines, where \Ha\ and \Hb\ refer only to 
the narrow component of the line. 
The main virtues of this technique, illustrated in Fig. \ref{fig:red}, are: 1) the lines are
relatively strong, 2) the line ratios are relatively insensitive to reddening corrections 
because of their close separation, and 3) at least at low redshift (z $\lesssim$ 0.5) the lines 
are accessible using ground-based optical telescopes.
Several samples have been selected in the past using the BPT diagrams and the method select 
AGN reliably with high completeness \citep{Zavadsky2000,Zakamska2003,Hao2005ss}.

At high redshift, however, the involved lines are redshifted out of the 
observed optical range and the classical BPT diagrams can no longer be used. In these  
circumstances, it is thus
desirable to devise a classification system that is based only on the blue part of the spectrum.

For this reason, \citet{Rola1997}, \citet{Lamareille2004}, and \citet{PerezMontero2007} proposed alternative 
diagrams based on the strong lines \OII, \NeIII, \Hb, and \OIIIb, which provide moderately effective 
discrimination between starbursts and AGN.
Since this technique is more recent than classical BPT diagrams, it has been used by fewer studies 
in the literature. We also note that, the use of the ratio of two lines that are not close to each 
other in wavelength (\lHb\  and \lOII) 
makes this diagram sensitive to reddening effects which, due to differential extinction of the 
emission lines and the stellar continuum \citep{Calzetti1994}, also affect the EW measurements. 

An important issue to address in AGN studies is their evolution.
The overall optical luminosity function of AGN, as well as that of different types of AGN, holds
important clues about the demographics of the AGN population, which in turn provides 
strong constraints on physical models and theories of AGN and galaxy co-evolution.

Many studies have been conducted and many results obtained in the past few years to 
constrain the optical luminosity function of  type--1 AGN at both low  
\citep{Boyle1988,Hewett1991,Pei1995,Boyle2000,Croom2004} and high redshift 
\citep{Warren1994, Kennefick1995,Schmidt1995,Fan2001,Wolf2003,Hunt2004,Bongiorno2007,Croom2009}. 
In contrast, there are not many type--2 AGN samples available in the literature and consequently very few studies of their 
evolution have been conducted. 

In the local Universe, \citet{Huchra1992} selected 25 Seyfert--1 and 23 Seyfert--2 galaxies from 
the CfA redshift survey \citep{Huchra1983} and used these AGN to measure their luminosity function.
\citet{Ulvestad2001} also computed the local luminosity function of a sample selected from the Revised 
Shapley-Ames Catalog \citep{Sandage1981}, and using the BPT diagrams, \citet{Hao2005lf} derived the 
luminosity function of a sample selected from the SDSS at z $<0.13$.\\ 
The only sample that spans a relatively wide redshift range, from the local Universe up to 
z$\sim$0.83, is that 
selected by \citet[][hereafter R08]{Reyes2008} from the SDSS sample, which is  
however limited to bright objects 
($10^{8.3}$ L$_{\odot} <$ L$_{\rm [OIII]} <$ 10$^{10} L_{\odot}$). Thus, 
a sample of type--2 AGN encompassing 
a wide redshift interval and including lower luminosity objects is highly desirable.

The zCOSMOS survey \citep{Lilly2007,Lilly2009} is a large redshift survey in the COSMOS field.
%that is being undertaken in 
%the COSMOS field with the VIsible Multi-Object Spectrograph (VIMOS) mounted on the ESO 8m 
%Very Large Telescope (VLT).  
From this sample, using the standard BPT diagrams at low redshift and the diagram from \citet{Lamareille2004} 
at high redshift, we selected a sample of 213 type--2 AGN in a wide redshift range (0.15$<z<$0.92) and 
luminosity range ($10^{5.5} \rm L_{\odot} < L_{\rm [OIII]} < 10^{9.1} \rm L_{\odot}$). Here we present the main 
properties of this sample, their \OIIIb\ line luminosity function, and the derived type--2 AGN fraction as a function of 
luminosity and redshift.

The paper is organized as follows: 
 Sect. \ref{sec:cosmos} presents a brief overview of the COSMOS project 
 and in particular of the zCOSMOS sample, while in Sect. \ref{sec:sample_selection} 
 we describe in detail the adopted method to select the sample.
 Sections \ref{sec:Opt_comp} and \ref{sec:Xcomp} compare our sample 
 with both other optical samples and with the X-ray selected sample in the same 
 field (XMM-COSMOS; \citealt{Hasinger2007,Cappelluti2009,Brusa2007}; Brusa et al., in prep) respectively.
 Finally, in Sect. \ref{sec:lf_method}  we derive our emission-line AGN 
 luminosity function, and in Sect. \ref{sec:lf_result}, we compare the results 
 with those in previous works and the derived evolutionary model, as well as 
 the type--2 AGN fraction as a function of luminosity and redshift.
 Finally, Sect. \ref{sec:conclusion} summarizes our work.
 
Throughout this paper, we use AB magnitudes and assume a cosmology with 
$\Omega_{\rm m}$ = 0.3, $\Omega_{\Lambda}$ = 0.7 and H$_{0}$ = 70 km s$^{-1}$ Mpc$^{-1}$.

\section{zCOSMOS observations and data processing} \label{sec:cosmos}

The Cosmic Evolution Survey \citep[COSMOS,][]{Scoville2007} is the largest HST survey 
(640 orbits) ever undertaken, which consists of imaging with the Advanced Camera 
for Surveys (ACS) of a $\sim$ 2 deg$^2$ field  with single-orbit I-band (F814W) exposures \citep{Koekemoer2007}.

COSMOS observations include the full and 
homogeneous coverage of the field with multi-band photometry: (i) UV with GALEX 
(Schiminovich et al., in prep.), (ii) optical multi-band data with CFHT and Subaru \citep{Capak2007}, 
(iii) near-infrared (NIR) with CTIO, KPNO \citep{Capak2007} and CFHT \citep{McCracken2009},  
(iv) mid-infrared (MIR) and far-infrared (FIR) with Spitzer \citep{Sanders2007}, 
(v) radio with VLA \citep{Schinnerer2007}, and (vi) X-rays with XMM and Chandra \citep{Hasinger2007,Elvis2009}. 

The zCOSMOS spectroscopic survey \citep{Lilly2007,Lilly2009} is a large redshift survey that is being undertaken in 
the COSMOS field using $\sim$ 600 hours of observations with  VIMOS
mounted on the ESO 8 m VLT. 
The survey has been designed to probe galaxy evolution and the effects of environment up to 
high redshift and to produce diagnostic information about galaxies and AGN.

The zCOSMOS spectroscopic survey consists of two parts: (1) zCOSMOS-bright 
is a pure-magnitude limited survey, which spectroscopically 
observes with the MR grism (R $\sim$ 600; 5550-9650 \AA) objects brighter than I=22.5. 
It will ultimately consist of spectra of about 20,000 galaxies selected across the entire COSMOS field. 
(2) In zCOSMOS-deep, sources are selected, within the central 1 deg$^2$, 
using color-selection criteria to cover the range 1.4 $< z <$ 3.0. In this case, 
observations are performed with the LR-blue grism (R $\sim$ 200; 3600-6800 \AA).

For both samples, spectra were reduced and spectrophotometrically calibrated using the VIMOS Interactive Pipeline Graphical Interface 
software \citep[VIPGI,][]{Scodeggio2005} and redshift measurements were performed with the help of an automatic 
package \citep[EZ,][]{Fumana2008} and then visually double-checked \citep[for more details, see][]{Lilly2007,Lilly2009}.  
Finally, line fluxes and equivalent widths (EWs) were measured using our automated pipeline 
platefit\_vimos (\citealt{Lamareille2009}; Lamareille et al., in prep), which simultaneously fits all the emission lines 
with Gaussian functions after removing the stellar continuum.

The results presented here are based on the first half of the zCOSMOS-bright survey which consists of 10,644 
spectra \citep[``10k sample'';][]{Lilly2008,Lilly2009}, corresponding to $\sim$ 33\% of the total number 
of galaxies in the parent photometric sample.
 
\section{The type--2 AGN sample} \label{sec:sample_selection}

We isolate a sample of type--2 AGN from the zCOSMOS bright sample, using the standard BPT 
(\OIIIb/\Hb\ versus \NII/\Ha\ and \OIIIb/\Hb\ versus \SII/\Ha), and the \lOIIIb/\Hb\ versus \OII/\Hb\ diagnostic diagrams. 

We first used the entire zCOSMOS 10k bright sample excluding duplicate objects, stars, and broad-line 
AGN. Our initial sample contained 8878 extragalactic sources and in particular 7010 
in the redshift range considered (0.15$\lsim$z$\lsim$0.45 and  0.5$\lsim$z$\lsim$ 0.92). 
We excluded the redshift range 0.45$<z<$0.5 because, for the wavelength range covered by the 
VIMOS MR grism, 
the lines \NII, \SII, and \Ha\ are redshifted outside the limit of the spectrum at z$\gsim$0.45 
and the \OII\ line enters the observed wavelength range only at z$\gsim$0.5.

Secondly, we applied a selection criterion based on the signal-to-noise ratio (S/N) of the 
lines %\footnote{defined as the ratio between 
%the segnal of the line and the noise of the continuum around the line.} 
involved in the considered diagnostic diagram. 
In particular, we selected only emission-line galaxies in the explored redshift range 
for which S/N(\OIIIb)$>$5 and the S/N of the other involved lines was S/N(other)$>$2.5. 
This criterion is based mainly on the \OIIIb\ line since (1) it is 
the only line always present in the observed wavelength range for our adopted redshift interval and (2) we use 
the \OIIIb\ line to compute the luminosity function, so higher quality is required for this line.\\
The sample extracted with this selection criterion consists of 3081 sources, which represents 44\% of the parent 
sample (7010 galaxies). Hereafter, we refer to this sample as the ``emission-line sample.''
%_________________________FIGURE__________________________________________
\begin{figure}                                                      
\begin{center}                                                      
\includegraphics[height=8.0cm,width=8.0cm]{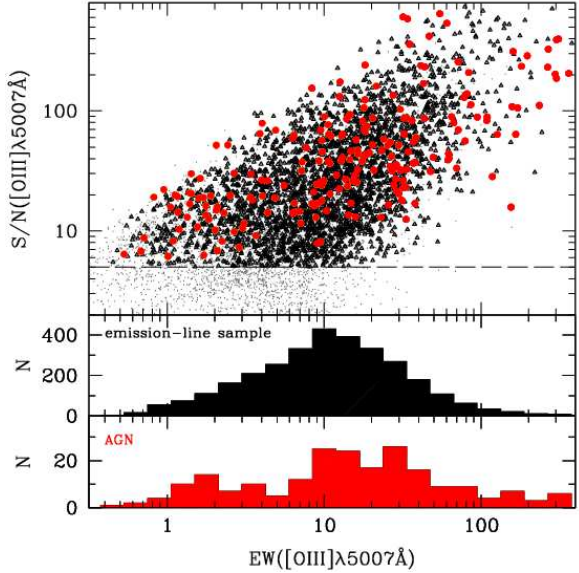} 
\caption{Observed EWs of the \OIIIb\ emission line versus the signal-to-noise ratio of the same line.
Grey points represent the parent sample of galaxies in the redshift range considered, 
while black triangles correspond to the emission-line sample (see text) obtained after applying 
our selection criteria (S/N(\OIIIb)$>$5 and S/N(oth)$>$2.5). The dashed line corresponds to the cut in 
S/N of the \OIIIb\ line.
Finally, red circles highlight the type--2 AGN sample selected on the basis of the line ratios. %(red for Sey--2 and 
%yellow for LINERs). 
The two bottom panels show the EW distribution of the emission-line 
sample and of the type--2 AGN sample, respectively.}
\label{fig:ew_sn}                                                     
\end{center}                                 
\end{figure}
Finally, we used the line ratios in the diagnostic diagrams to classify the selected galaxies  
into two main classes (star-forming galaxies and type--2 AGN), drawing a sample of 213 type--2 AGN. 
%(red circles in figure \ref{fig:ew_sn}). 
%This work uses line fluxes and equivalent widths measured by our automated pipeline 
%(platefit\_vimos; \citealt{Lamareille2009}; Lamareille et al., in prep) which simultaneously fits all the emission lines 
%with Gaussian functions after removing the stellar component.

The total parent sample, the emission-line sample, and the AGN sample are shown in Fig. \ref{fig:ew_sn}, 
respectively, as grey points, black triangles, and red circles. %\textbf{filled circles (red for Sey--2 and yellow for LINERs).}
This plot shows the observed equivalent widths (EW) of the \OIIIb\ emission line 
versus the S/N of the same line, highlighting the adopted cut in \OIIIb\ S/N (dashed line).
%\textbf{It is interesting to note that as in local universe \citep{xxx}, LINERs have lower \OIIIb\ EWs compared 
%to the Seyfert--2 sample.}
%In the same plot red circles show the position of the type--2 AGN selected on the basis of the line ratios.\\ 
Moreover, the bottom two panels show the EW distribution of the final 
emission-line sample and that of the type--2 AGN sample.

\begin{figure*}                                                      
\begin{center}                                                      
\includegraphics[height=12.0cm,width=15.0cm]{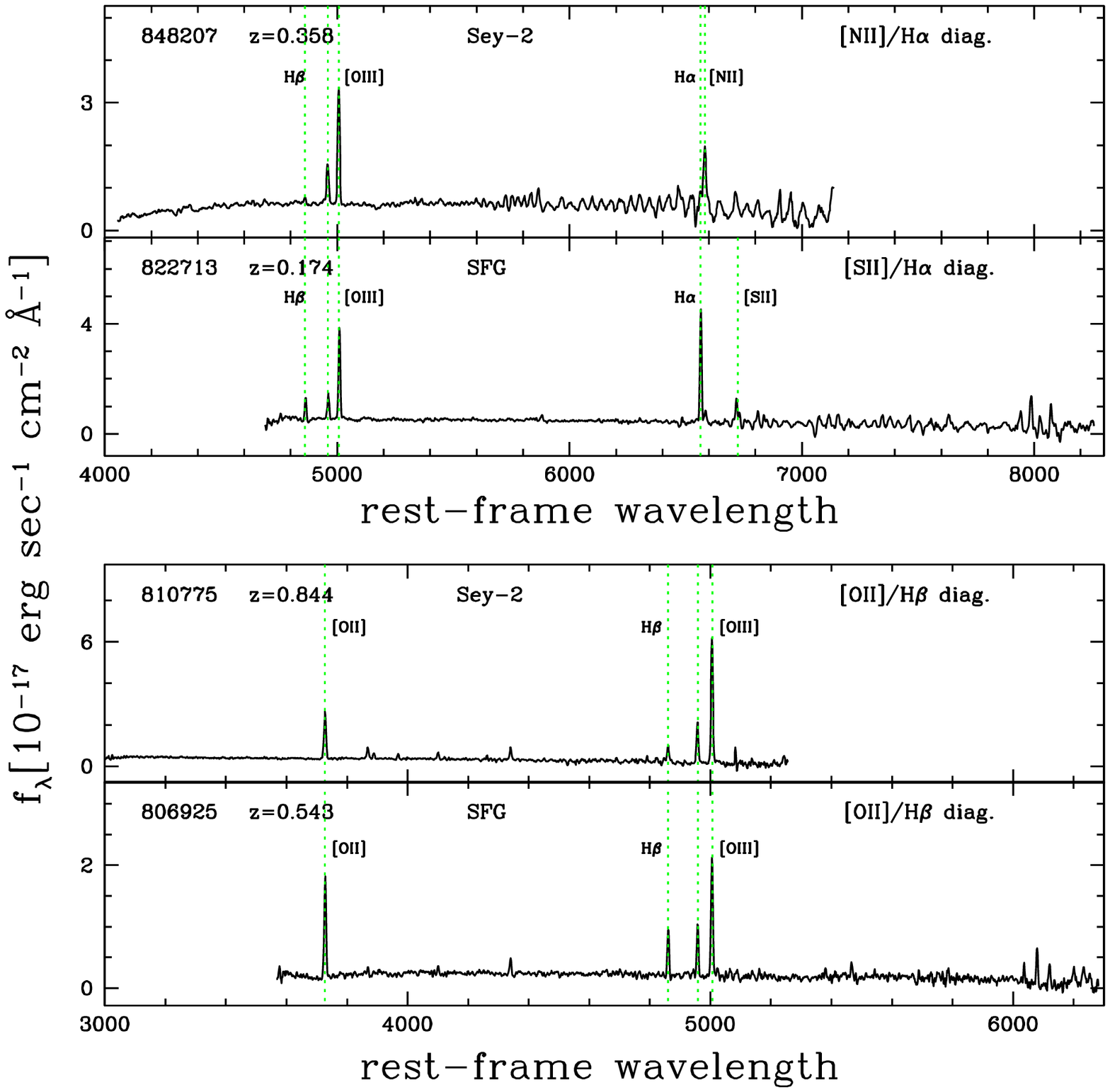}
\\
\includegraphics[height=12.0cm,width=15.0cm]{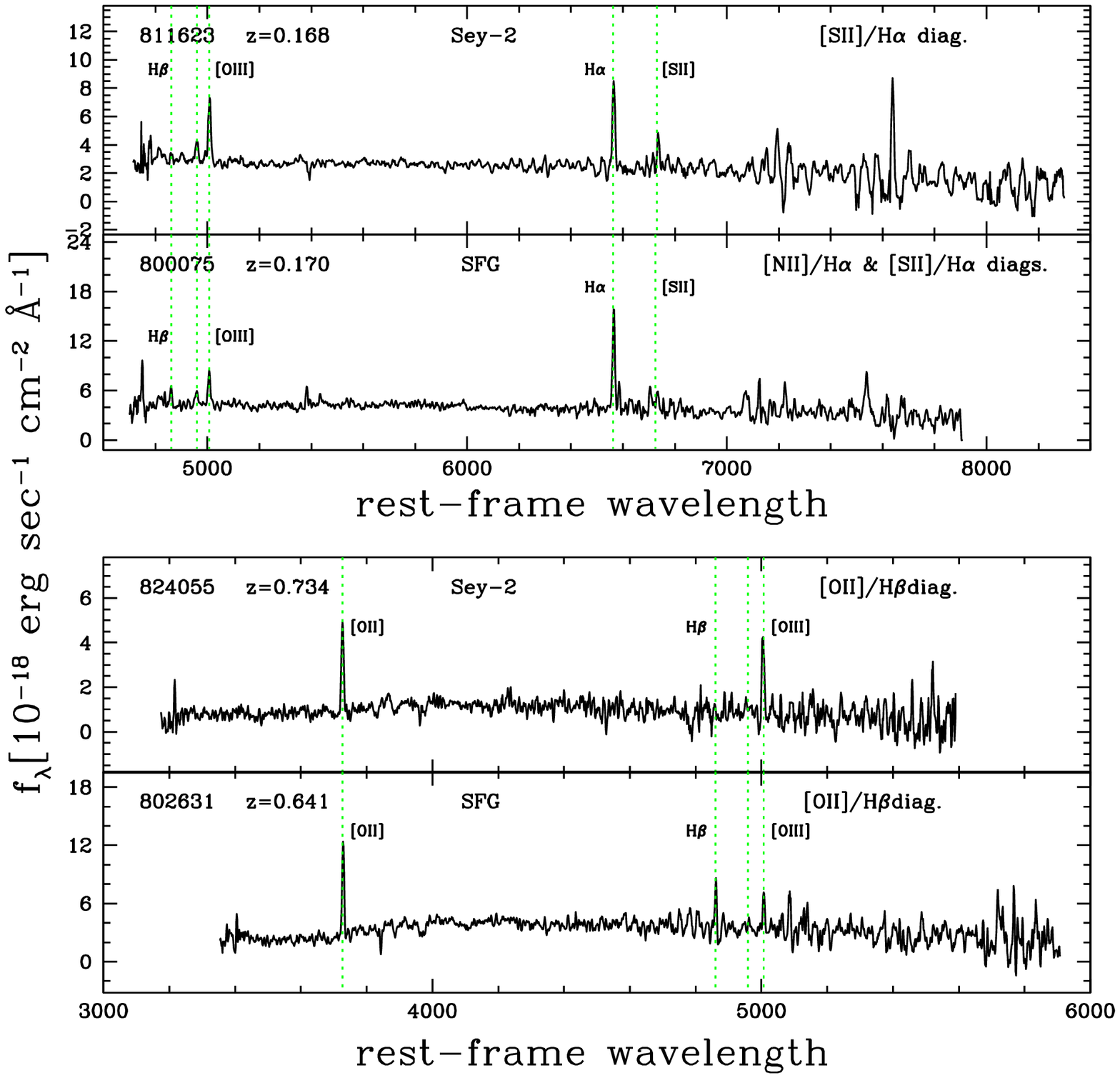}
\caption{Examples of zCOSMOS spectra, smoothed by 5 pixels. \textit{Upper panels:} 
Four examples of rest-frame spectra with higher \OIIIb\ S/N ($\gsim$ 150) of objects classified as Sey--2 and star-forming galaxy (SFG) 
in the low and high redshift bin, respectively.
\textit{Lower panels:} the same as above but showing spectra with lower \OIIIb\ S/N (20$\lsim$S/N$\lsim$70).}
\label{fig:spectra_examples}                                                     
\end{center}                                 
\end{figure*}

Figure \ref{fig:spectra_examples} shows some representative zCOSMOS spectra that fulfill these criteria. 
The upper and lower panels correspond, respectively, to higher ($\gsim$ 150) and lower (20$\lsim$S/N$\lsim$70) \OIIIb\ S/N.
In both panels, we show 4 examples of rest-frame spectra of Sey--2 and SFGs at different 
redshifts, two of them corresponding to the low-redshift bin  
(0.15 $<z<$ 0.45; see Sect. \ref{sec:sel_lowz}) and the other two to the high redshift bin 
(0.50 $<z<$ 0.92; see Sect. \ref{sec:sel_highz}).

In the following sections, we discuss in detail the type--2 AGN selection procedure in the 
two redshift intervals.

\subsection{Selection at 0.15 $<z<$ 0.45} \label{sec:sel_lowz}

In this redshift range, we used two diagrams based on
line-intensity ratios constructed from \lHb, \lOIIIb, \lHa, \lNII, and \lSII. 
In particular, we used the standard BPT diagrams proposed by \citet{Baldwin1981} 
and revised by  \citet{Veilleux1987}, which consider the plane  \OIIIb/\Hb\
vs \NII/\Ha\  (hereafter \reda\ diagram) and \OIIIb/\Hb\ versus \SII/\Ha\  (hereafter \redb\ diagram). 
When both \NII\ and \SII\ lines are measured with S/N$>$2.5, the classification was derived 
by combining the results obtained from both diagrams.
%Due to the metallicity sensitivity of \NII/\Ha, the \reda\ diagram should be more sensitive to the 
%presence of low-level AGN than red2. However, in our analysis
%we considered both diagrams and we derived the classification combining the results obtained by both diagrams.
%For this reason the classification obtained from 
%the \textit{\reda\} BPT diagram (which is based on \NII/\Ha\ line ratio) is more reliable and will 
%be used through this paper as reference for the \textit{red2} diagram.
 
\begin{figure*}                                                      
\begin{center}
\subfigure[]{                                                      
	\includegraphics[height=8.0cm,width=8.0cm]{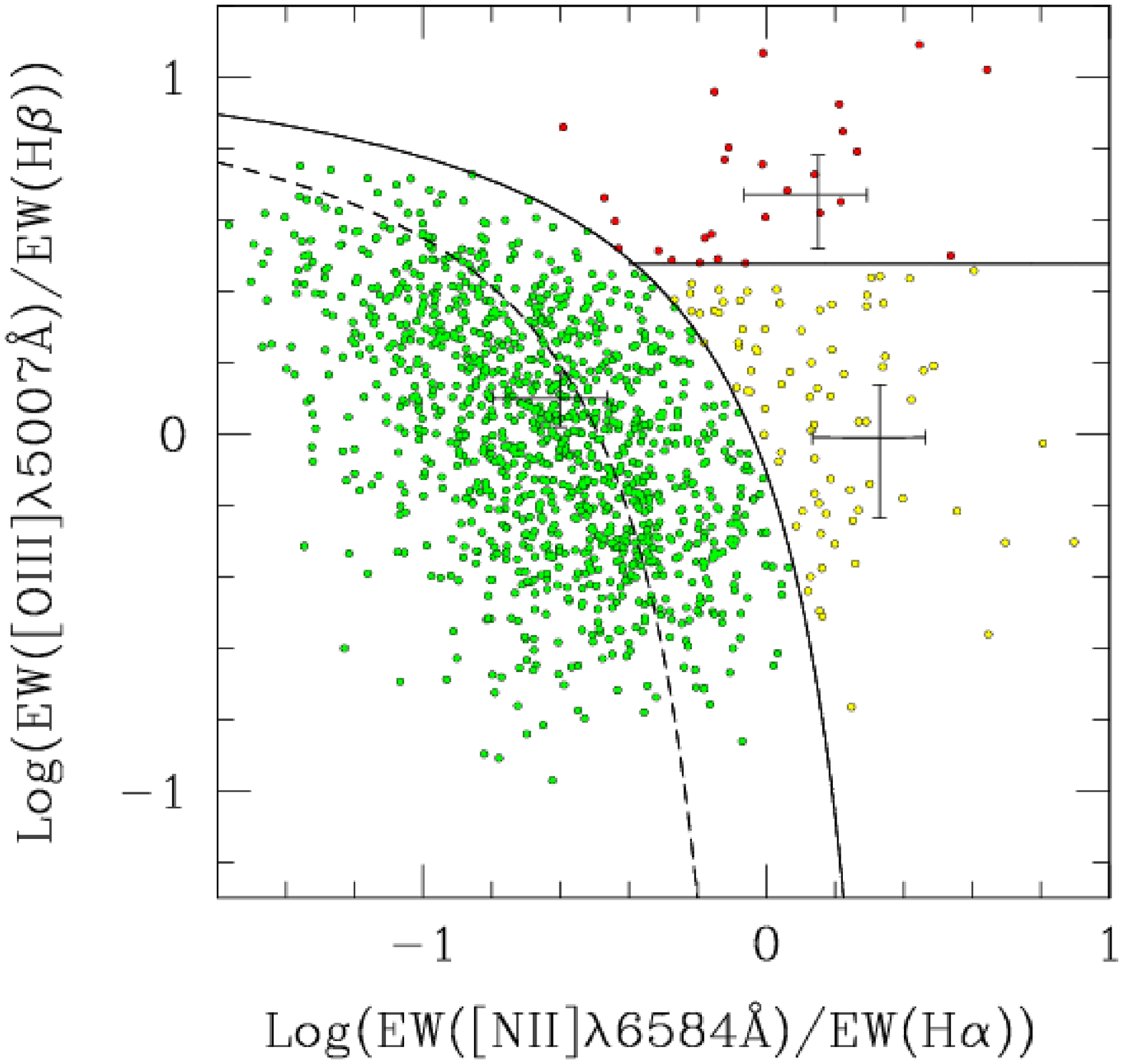}
}
\subfigure[]{	
        \includegraphics[height=8.0cm,width=8.0cm]{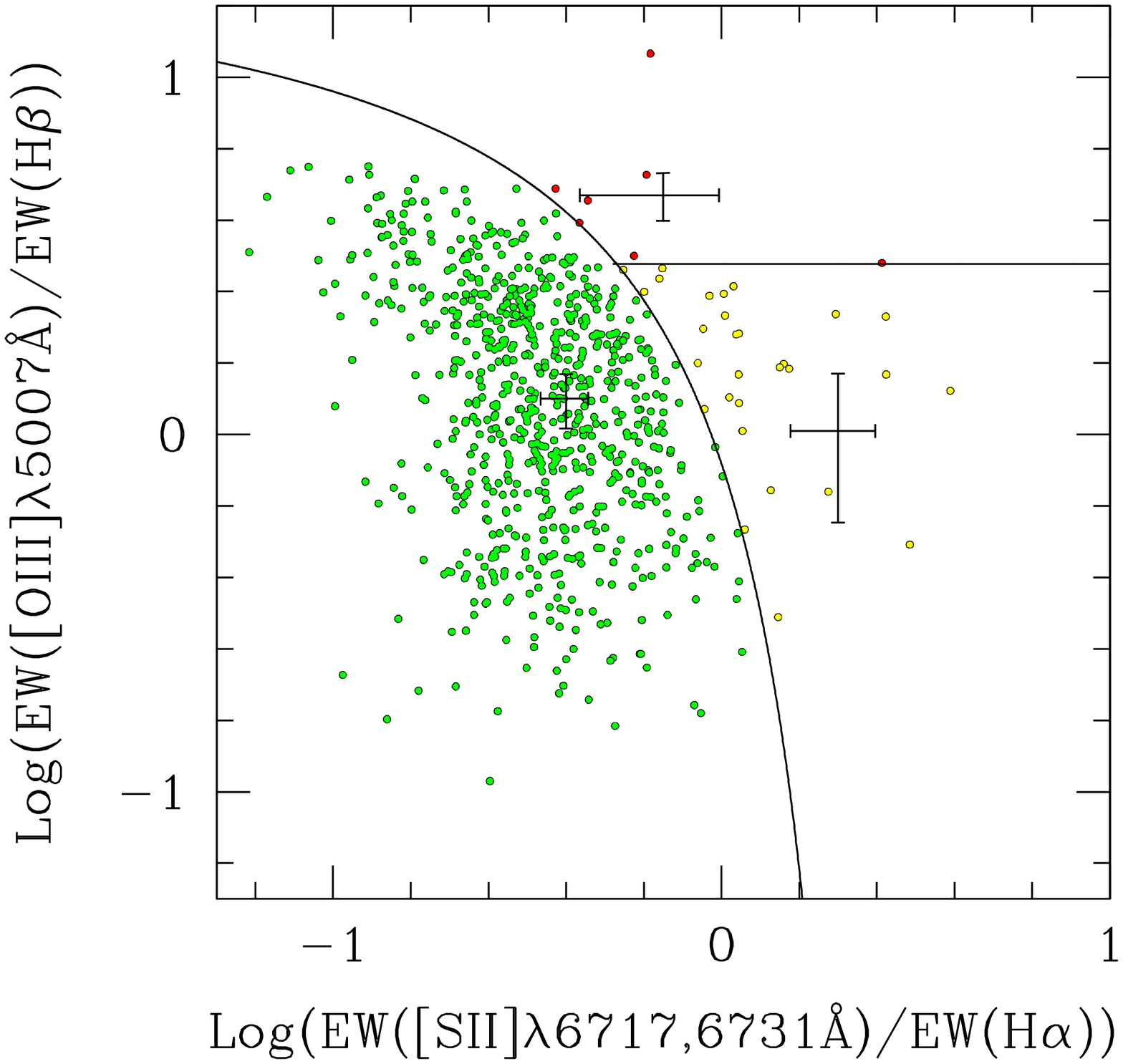}
}
\caption{\textbf{[0.15 $<z<$ 0.45]:} \textit{(a)} log(\OIIIb/\Hb) versus log(\NII/\Ha) (\reda\ diagram) and 
\textit{(b)} log(\OIIIb/\Hb) versus log(\SII/\Ha) (\redb\ diagram) BPT diagrams. 
The solid lines show the demarcation between SFG and AGN used in this 
work, which corresponds to the one defined by \citet{Kewley2001} in panel (a) and a modified version in panel (b), obtained 
by adding a 0.15 dex in both axes (see text). The 
dashed line in panel (a) shows as a comparison the demarcation defined by \citet{Kauffmann2003}. 
Composite objects are expected to be between this line and the \citet{Kewley2001} one.
A typical errorbar for SFG, Sey--2, and LINER is also shown in both diagrams.
For spectra where both \NII\ and \SII\ lines satisfy the selection criteria (S/N$>$2.5), the classification is performed using 
both diagrams (see Sect. \ref{sec:sel_lowz}): 
red circles correspond to objects classified as Sey--2, yellow circles correspond to the objects 
classified as LINER, while green circles are objects classified as SFG.}
\label{fig:red}                                                     
\end{center}                                 
\end{figure*}

The exact demarcation between star-forming galaxies and AGN in the BPT diagrams is subject to 
considerable uncertainty. In this redshift bin, we assumed the theoretical upper limits 
to the location of star-forming galaxies in the BPT diagrams derived by \citet{Kewley2001}. 
%using a combination of photoionization and stellar population synthesis models. 
However, following \citet{Lamareille2004}, we added a 0.15 dex shift to both axes 
to the separation line in the \redb\ diagram (Eq. \ref{eq:red2}) to 
obtain closer agreement between the classifications obtained with the two diagrams.
Using the standard division line \citep[without the 0.15 
dex shift; see Eq. (6) of][]{Kewley2001}, the disagreement between the \reda\ and \redb\ 
classifications would be 25\%, significantly higher than the 5.5\% obtained by adding this 0.15 
dex shift (see below). 
Moreover, for consistency with the selection at 0.5 $<z<$ 0.92 (see Sect. \ref{sec:sel_highz}), 
EWs were used instead of fluxes. Since the wavelength separation between the emission 
lines involved in these diagrams is small, the use of either EWs or fluxes 
as diagnostics is largely equivalent and produces very similar results.
%This choice is motivated by the fact that we want to minimize 
%the contamination in our AGN sample from SFG.
The analytical expressions we adopted for the demarcation curves between starburst and AGN-dominated objects 
are the following

\begin{equation}
\log\left(\frac{\rm \OIIIb}{\rm H\beta}\right)=\frac{0.61}{\log\left(\rm \NII/\rm H\alpha\right)-0.47}+1.19;
\label{eq:red1}
\end{equation}

\begin{equation}
\log\left(\frac{\rm \OIIIb}{\rm H\beta}\right)=\frac{0.72}{\log\left(\rm \SII/\rm H\alpha\right)-0.47}+1.45.
\label{eq:red2}
\end{equation}
Starburst galaxies are located below these lines, while type--2 AGN are above (see solid lines in Fig. \ref{fig:red}). 
In Fig. \ref{fig:red} panel (a), we also show (dashed line) the 
 demarcation line defined by \citet{Kauffmann2003}. The intermediate region in-between this line and the 
\citet{Kewley2001} division line is the parameter space where composite objects are expected. 

%____________________FIGURE_________________________________________

%-----------------------------

%____________________FIGURE_________________________________________

%-----------------------------

In the region of type--2 AGN, we can distinguish further between Seyfert--2 galaxies and low ionization nuclear 
emission regions \citep[LINERs;][]{Heckman1980}.
We applied the separation limit based on the \lOIIIb/\lHb\ ratio 
\citep[\OIIIb/\Hb $<$ 3.0 for LINERs;][]{Ho1997class}. 
It is still unclear whether all LINERs are AGN. Many studies have been conducted in different passbands 
to understand the nature of these objects. In the UV band, \citet{Barth1998} and \citet{Maoz2005} found nuclear emission in $\sim$25\% 
of the observed LINERs. Moreover, about half of them appear point-like at the resolution of HST, thus being candidate AGN. 
In the radio band, \citet{Nagar2000} found that $\sim$50\% of LINERs have a compact radio core. Subsequent 
studies \citep{Falcke2000} confirmed the existence of compact, high-brightness-temperature cores, suggesting that 
an AGN is responsible for the 
radio emission rather than a starburst. In the optical band, \citet{Kewley2006}, studying the 
host properties of a sample of emission-line galaxies selected from the SDSS, found that 
LINERs and Seyfert galaxies form a continuous sequence in L/L$_{\rm EDD}$, thus suggesting 
that the majority of LINERs are AGN. 
Finally, from the X-ray band \citet{Ho2001liner}, studying the X-ray properties 
of a sample of low-luminosity AGNs, 
found that at least 60\% of LINERs contain AGNs, consistent with the estimates of \citet{Ho1999liner}.
The same percentage was also recently found by \citet{Gonzalez2006} for a sample of 
bright LINER sources. 
In the computation of the luminosity function, we consider the total sample of type--2 AGN 
and LINERs without any distinction.

%However, in the computation of the luminosity function, we consider the total sample of type--2 AGN 
%and LINERs without any distinction. 
%Recent studies of local samples of LINERs have shown in fact 
%that most of the LINERs (at least 60\%) host a low luminosity AGN in their nuclei \citep{Gonzalez2006}.

%Recently, \cite{Kewley2006} proposed a new empirical classification scheme to separate 
%Sey--2 and LINERs in the \textit{red2} diagram.
%The main difference from the \cite{Heckman1980} method is that it is not based only on the 
%\OIIIb/\Hb\ ratio but also on the \SII/\Ha\ ratio (see the green line in figure \ref{fig:red}).
%We tested this new classification in the \textit{red2} diagram but we found that 6 of the sources 
%classified as Seyfert--2 on the basis of the \textit{red1} diagram would be classified as LINER 
%according to the new classification. 
%On the contrary, the old classification by \cite{Heckman1980} would define them Seyfert--2, in agreement 
%with the \textit{red1} classification.

The zCOSMOS sample in the 0.15$<z<$0.45 redshift range consists of 2951 sources of which 1461 satisfy 
our emission-line selection criteria as shown in the diagnostic planes.
Many of them were classified in only one of the two diagrams, but 614 objects have both \NII\ and 
\SII\ lines measured and were thus classified using both diagrams. In these cases, the classification 
was performed on the basis of the position of the objects in both diagrams. 
For 580 of them (94.5\%), the two classifications were consistent with each other, while 
the remaining 5.5\% of the objects were classified differently using the two diagrams.
We confirmed that all these objects are, in at least one of the two diagrams, close to the separation line, 
where the classification is not secure. For this reason, we classified these objects on the basis of their distance 
from the division line in the diagram.
In particular, a classification is taken as the most likely solution if its distance (normalized to its error) from the 
demarcation line is the greatest of the two solutions. Since the two diagnostic diagrams have the 
same y-axis, the distance is computed along the x-axis. Using this method, 27/34 ($\sim$ 80\%) objects were  
classified according to the \reda\ diagram, and the remaining 20\% using the \redb\ diagram.

The final type--2 AGN sample extracted in this redshift range consists of 128 sources out of a total sample of 
1461 sources. Thirty-one of them are classified as Seyfert--2 and 97 as LINERs (see Table \ref{tab:statistic}).
In this redshift range, LINERs constitute $\sim$75\% of the AGN sample. As comparison, 
the fraction of LINERs found by \citet{Lamareille2009} in the same redshift range, 
using the data from the Vimos-VLT Deep Survey (VVDS), is $\sim$55\% for the wide sample (I$_{\rm AB}<22.5$) and 
$\sim$66\% for the deep one (I$_{\rm AB}<24.0$).    
The lower percentages in the VVDS sample are not surprising. Given the lower resolution of VVDS spectra 
compared to zCOSMOS, there are more difficulties in deblending the \Ha\ and \NII\ lines and this is particularly 
true for objects where these two lines have similar fluxes, as LINERs.

%__________________FIGURE__________________________
\begin{figure}                                                      
\begin{center}                                                      
\includegraphics[height=8.0cm,width=8.0cm]{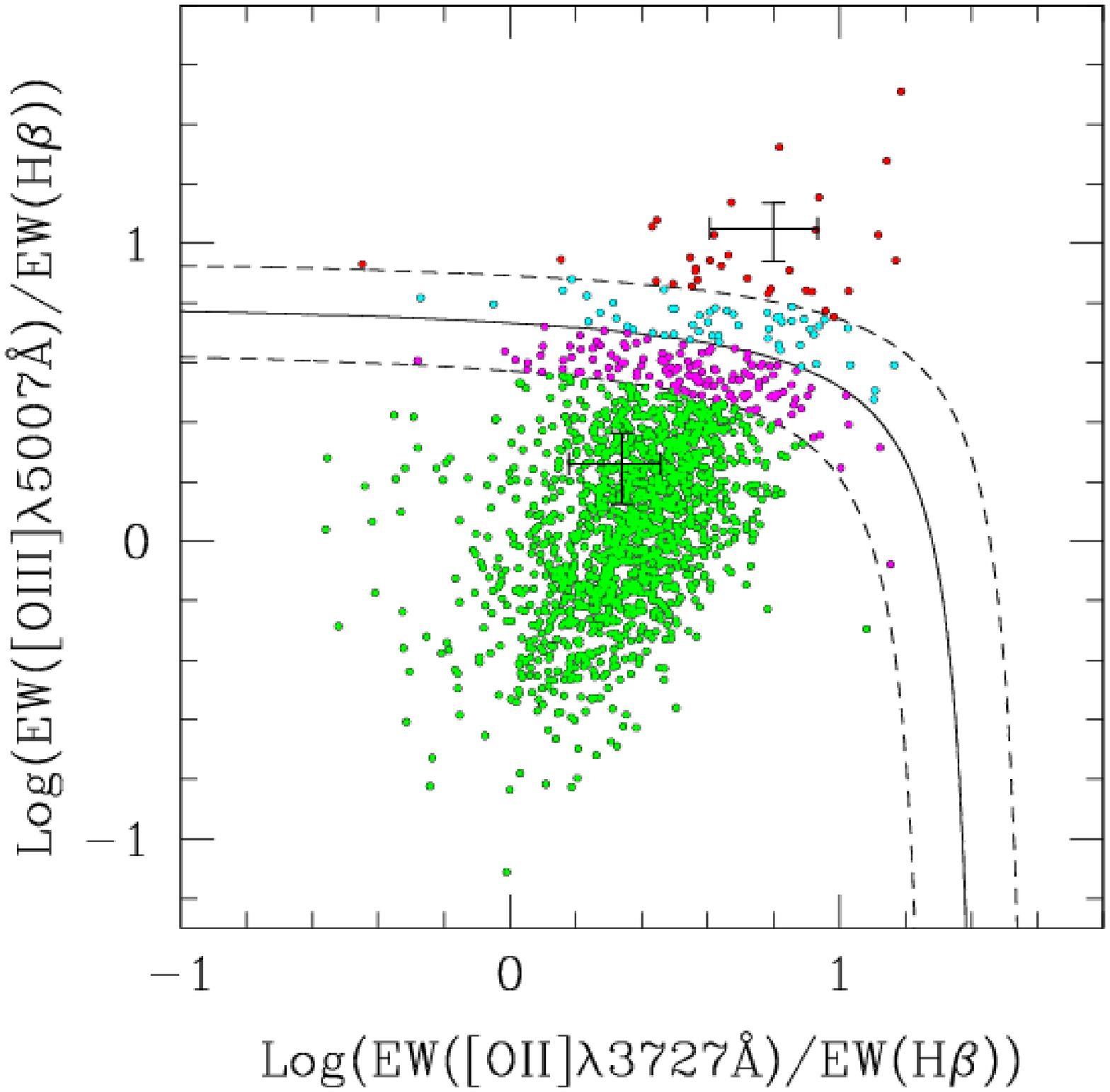}
\caption{\textbf{[0.5 $<z<$ 0.92]:} log(\OIIIb/\Hb) versus log(\OII/\Hb) (\blue\ diagram). 
The solid line shows the demarcation between SFG and AGN defined by \citet{Lamareille2004} and 
used in this work. Moreover, the dashed lines define an intermediate region close to the 
demarcation line where intermediate objects, i.e., candidate Seyfert--2 and candidate SFGs, 
are expected to be found.
Red circles correspond to objects classified as Sey-2 and
 green circles to SFG. 
 Cyan circles and magenta circles are candidate Sey--2 
 and candidate SFG, respectively. A typical errorbar for SFG and AGN is also shown.}
\label{fig:blue}                                                     
\end{center}                                 
\end{figure}
%--------------------------------------------------
\begin{table*}
\begin{center}
\begin{tabular}{c |c c c |c |c c |c }
\hline
\hline
\multicolumn{8}{c}{Analyzed sample = \textbf{3081}  galaxies satisfying the criteria S/N(\OIIIb)$>$5 \& S/N(oth)$>$2.5 }\\
\hline
&\multicolumn{3}{|c|}{0.15 $<z<$ 0.45}&0.50 $<z<$ 0.92 & & &\\
&\multicolumn{3}{|c|}{N=1461}&N=1620 & & &\\
\hline

class&\reda\ only&\redb\ only&both&\blue& TOT & X-ray det. & X-ray fraction\\

\hline
\textbf{Sey--2}  &    24(6)&      4  &    3 (1)   &    32(6) &  63 &  13  &    20\%\\
\textbf{LINER}   &     70(3) &   17   &  10(1)   &   --   &  97 &  4  &    4\%\\
\textbf{Sey--2 cand}  &  --   &   --  &    --   &    53(6)  &  53  & 6 &  11.3\%\\
SFG    &   603 (6) &   163   &  567(6) &   1398(25)   & 2731 &  37  &  1.3\% \\
SFG cand   &   --  &    --  &    -- & 137(6)   &     137 & 6   &   4.4\%\\
\hline
\hline

\end{tabular}
\end{center}
\caption{For each class of objects, the table shows the number of objects selected according 
to the different diagnostic diagrams (\reda, \redb, and \blue) and the fraction of them 
showing X-ray emission (numbers in parenthesis).
The last three columns instead show the total numbers and the number and fraction of X-ray detected objects 
for each class.}
\label{tab:statistic}
\end{table*}

\subsection{Selection at 0.50 $<z<$ 0.92} \label{sec:sel_highz}

In this redshift range, we used the diagnostic diagram originally proposed by \citet{Rola1997} and 
later analyzed in detail by \citet{Lamareille2004}, i.e., 
\lOIIIb/\Hb\ versus \OII/\Hb\ (hereafter \blue). The separation in this diagnostic diagram 
was derived empirically, on the basis of the 2dFGRS data, 
 by studying the position in this diagram of AGN and star-forming galaxies for which a previous 
classification based on the standard red diagrams was available.
The analytical expression defined in terms of EW for the demarcation curves between starburst galaxies 
and AGN is

\begin{equation}
\log\left(\frac{\rm EW(\OIIIb)}{\rm EW(H\beta)}\right)=\frac{0.14}{\log\left(\rm EW(\OII)/\rm EW(H\beta)\right) - 1.45}+0.83
\end{equation}
This diagram allows us to distinguish between Seyfert--2 galaxies and star-forming galaxies.
Moreover, following \citet{Lamareille2004} it is also possible to define an 
intermediate region close to the demarcation line (dashed lines in Fig. \ref{fig:blue}). 
Intermediate objects, i.e., candidate Seyfert--2 and candidate SFGs, are expected to lie in this region.

Since this diagram uses the ratio of two lines that are not close in wavelength,  
it is sensitive to reddening effects. The use of EWs instead of fluxes removes a direct dependence 
on reddening. However, since the reddening affects in a different way the emission lines  
and the underlying stellar continuum, it influences the ratio \OII/\Hb\ \citep{Calzetti1994}.
The final type--2 AGN sample extracted in this redshift range consists of 85 sources out of a total sample of 
1620 sources that satisfy our selection criteria.
Thirty-two of them are classified as Seyfert--2, and 53 as candidate Seyfert--2 galaxies 
(see Table \ref{tab:statistic}).

\subsection{The final type--2 AGN sample}

Summarizing, our final type--2 AGN sample consists of 213 objects out of a total sample of 
3081 galaxies with S/N(\OIIIb)$>$5 and S/N(oth)$>$2.5 in the redshift ranges 0.15$<z<$0.45 and 0.5$<z<$0.92.
Star-forming galaxies, which lie below the curves, represent $\sim$ 93\% of the 
sample, while type--2 AGN constitute only $\sim$ 7\% of the studied sample. 
In particular, 63 of the AGN are Sey--2, 53 %($\sim$25\%) 
are Sey--2 candidates (they lie in an intermediate region in the \blue\ diagram), and 
97 %($\sim$46\%) 
are LINERs selected from the \reda\ and \redb\ diagrams. No LINERs were selected from the \blue\ 
diagram. We discuss the number of possible LINERs missed in this diagram in Section \ref{sec:Xcomp}.
Given the luminosity range covered by our sample, contamination 
from narrow-line Sey--1 is expected to be of the order of few percent 
\citep[see e.g.,][]{Zhou2006}.

Figures \ref{fig:red} and \ref{fig:blue} show the position of sources in the three 
diagnostic diagrams used to classify them. Table \ref{tab:statistic} indicates 
the number of objects selected from each diagram, while the full catalog, containing position, 
redshift, I$_{\rm AB}$ magnitude, 
\OIIIb\ luminosity, and classification, can be found in Table \ref{tab:catalog}.

%__________________FIGURE__________________________

\begin{figure}                                                      
\begin{center}                                                      
\includegraphics[height=8.0cm,width=8.0cm]{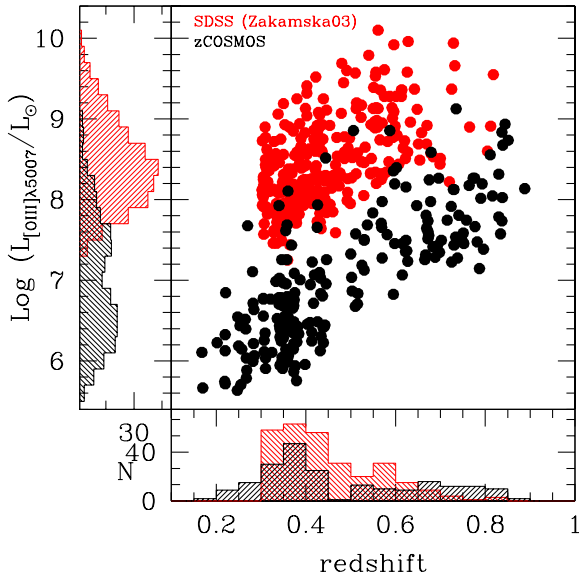} 
\caption{Redshift and \OIIIb\ luminosity distribution of the zCOSMOS type--2 AGN sample (black) 
compared to the SDSS sample (red) selected by \citet{Zakamska2003}. While the redshift ranges are very similar, 
the luminosity ranges covered by the two samples are complementary.}
\label{fig:comp_sdss}                                                     
\end{center}                                 
\end{figure}
%--------------------------------------------------

\section{Comparison with other optical surveys} \label{sec:Opt_comp}

As discussed above, type--2 AGN have similar spectral continua to normal star-forming
galaxies and hence their optical selection is challenging.
The zCOSMOS spectra allowed us to select a sample (see Fig. \ref{fig:comp_sdss}) that spans a 
wide range in both redshift  (0.15$<z<$0.92) and luminosity 
($10^{5.5} L_{\odot}<L_{\rm \OIIIb}<10^{9.1}  L_{\odot}$).
  
The only other sample that spans a comparable redshift range is selected in a very similar way
from the Sloan Digital Sky Survey (SDSS) Data Release 1 (DR1) by \citet{Zakamska2003}.
Their sample consists of 291 type--2 AGN at 0.3$<z<$0.83 (the redshift range chosen 
to ensure that the \lOIIIb\ line is present in all spectra). 
However, as shown in Fig. \ref{fig:comp_sdss}, this sample is significantly brighter than 
the zCOSMOS sample, spanning the luminosity range $10^{7.3} L_{\odot}<L_{\rm \OIIIb}<10^{10.1}L_{\odot}$.
 
From a three times larger SDSS catalog, combining different selection methods, R08 derived the 
luminosity function of a larger sample of type--2 AGN (887 objects within $\sim$ 6293 deg$^{2}$) with  
z$<0.83$ and a higher lower limit to its \OIIIb\ luminosity 
($10^{8.3} L_{\odot} < L_{\rm [OIII]} < 10^{10} L_{\odot}$) than the original SDSS 
sample from \citet{Zakamska2003}. With almost the same redshift range as the zCOSMOS sample, but at 
brighter luminosities, the SDSS sample of R08 complements our sample constraining 
the bright end of the luminosity function (see Fig. \ref{fig:lf} and Sect. \ref{sec:lf_result}).

%___________________________FIGURE___________________________________
\begin{figure*}                                                      
\begin{center}   
\subfigure[]{                                                   
      \includegraphics[height=8.0cm,width=8.0cm]{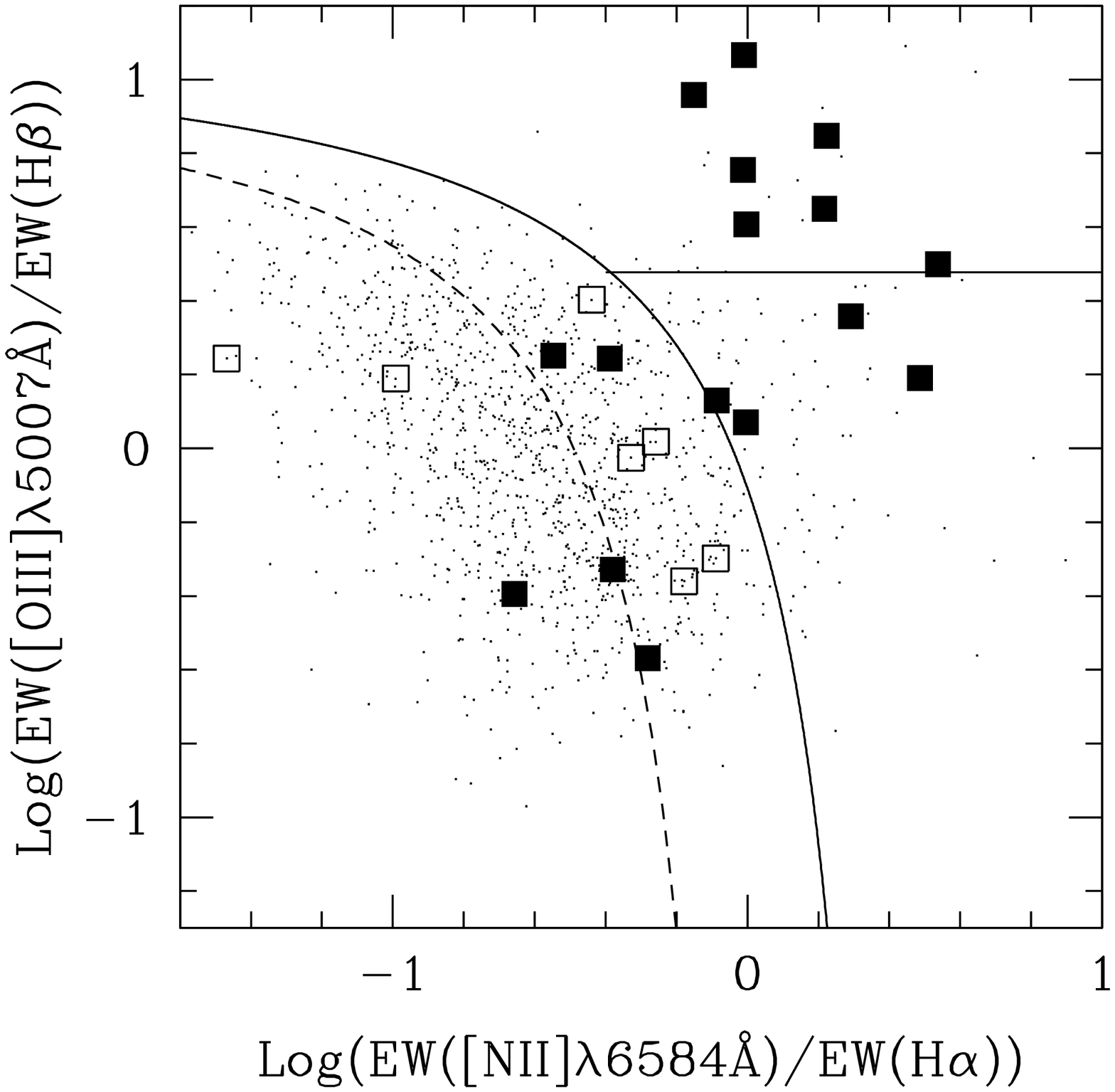}
      }
%\subfigure[]{
%	\includegraphics[height=5.0cm,width=5.0cm]{figure/red2_X.ps}
%}
\subfigure[]{
	\includegraphics[height=8.0cm,width=8.0cm]{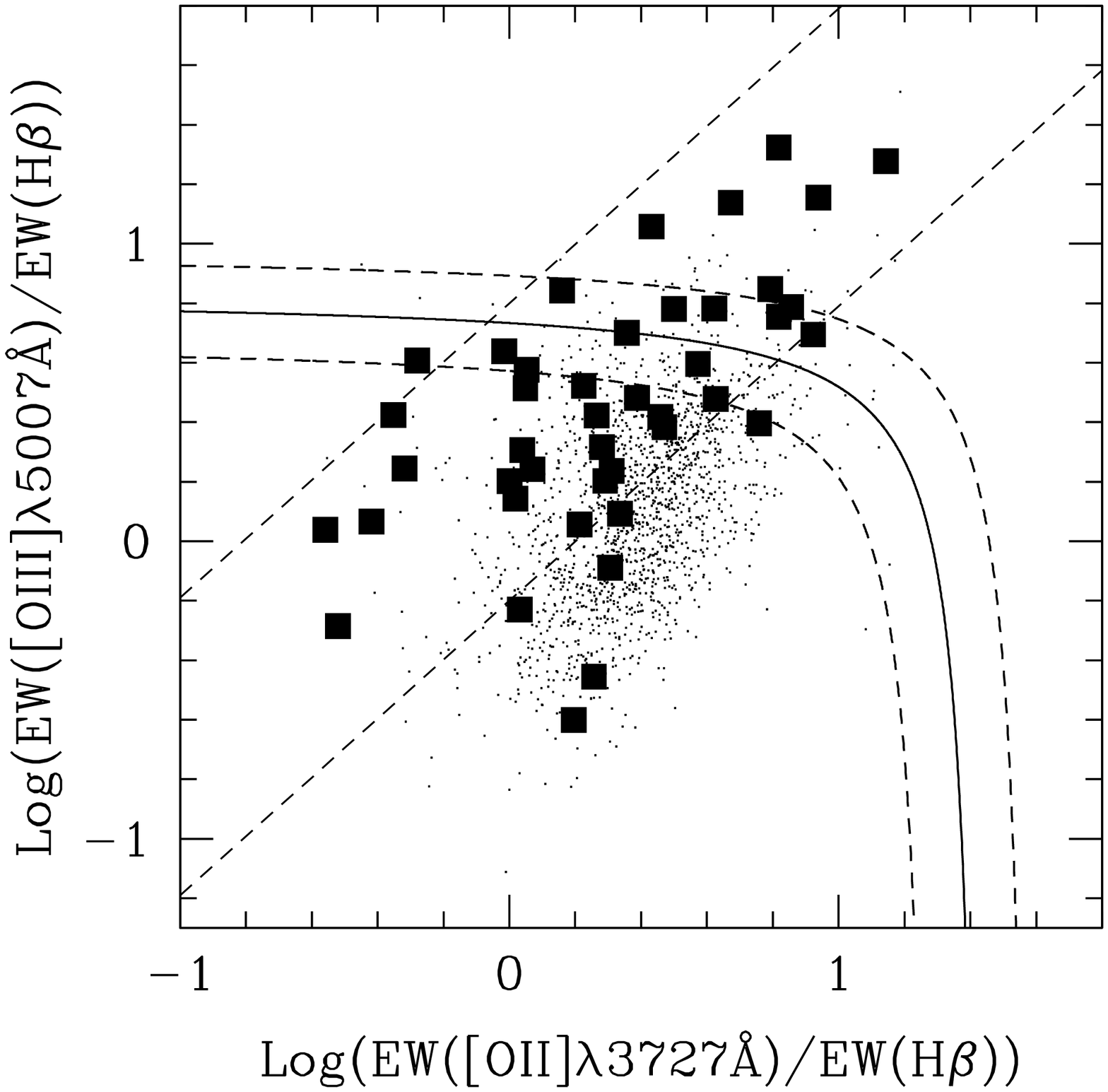}
}
\caption{As in Figs. \ref{fig:red}(a) and \ref{fig:blue}, but now showing the positions of X-ray 
detected sources. Open and filled symbols correspond to different X-ray luminosities 
(open for L$_x<10^{42}$ erg s$^{-1}$, and filled for L$_x>10^{42}$ erg s$^{-1}$).
%Moreover, in panel (b) different symbols refer to the measured N$_{\rm H}$ of the sources.
%Upward pointing triangles for N$_{\rm H}>$10$^{22}$ cm$^{-2}$, 
%downward pointing triangles for N$_{\rm H}<$10$^{22}$ cm$^{-2}$, and squares when N$_{\rm H}$ is not available.   
The two dashed lines in the same panel mark the preferred region occupied by X-ray detected objects.}
 \label{fig:all_X}                                                     
\end{center}                                 
\end{figure*}

\section{Comparison with the X-ray sample} \label{sec:Xcomp}

%(Hasinger et al. 2007, Cappelluti et al. 2007, 2008, Brusa et al. 2007; 2008)
Of our total analyzed sample, 66/3081 galaxies (2.1\%) have an X-ray counterpart (XMM catalog, \citealt{Cappelluti2009}, \citealt{Brusa2007}, Brusa et al., in prep.). 
Twenty-three of them are optically classified as AGN, while 43 of them are classified as SFG (see Table \ref{tab:statistic}). 

In Fig. \ref{fig:all_X}, we show the two main diagnostic diagrams (\reda\ and \blue), where X-ray sources are 
indicated with different symbols depending on their X-ray luminosity: 
open for L$_{[2-10]\rm keV}<10^{42}$ erg s$^{-1}$ and filled for L$_{[2-10]\rm keV}>10^{42}$ erg s$^{-1}$, 
which is the classical limit taken to define a source as an AGN \citep{Moran1999}.

The standard red diagnostic diagram (low redshift, panel (a)) broadly agrees with the X-ray classification.
All X-ray sources with  L$_{\rm x}<10^{42}$ erg s$^{-1}$ (6 objects) are in the SFG locus, while most
 (11 sources) of the luminous X-ray sources are indeed in the AGN locus.
There are 5 sources in the SFG region of the \reda\ diagram with  L$_{\rm x}>10^{42}$ erg s$^{-1}$. 
Two of them lie close to the division line and can indeed be explained by considering the errors 
in the EW line measurements. Moreover, these two sources lie above the \citet{Kauffmann2003} division line, 
where composite (SF+AGN) objects are expected to be.
Two more objects are just on top of the \citet{Kauffmann2003} division line and the remaining object is 
located fully in the SF region. Closer examination of the spectra of the latter three objects confirms their 
optical classification as SF galaxies and does not reveal any feature characteristic of AGN that 
would explain their high X-ray emission.
We can conclude that in the red diagnostic diagrams the optical and the X-ray classification agree 
at the (75-85)\% level: 18-20 sources out of 23 have the same classification, four are border-line cases and 
one object is clearly misclassified.

In the \blue\ diagnostic diagram, in contrast, the situation is far less clear.
%It is interesting to note that while the standard diagnostics seem to agree with the X-ray
%classification, with most of the luminous X-ray sources being indeed in
%the AGN locus, the situation is much less clear for the blue diagram.
In particular, we found that 31 out of 43 (72\%) X-ray sources 
with L$_{2-10\rm keV}>10^{42}$ erg s$^{-1}$ fall in the region of star-forming galaxies. 
Moreover, the position of most of them and of almost all the brightest sources 
(L$_{\rm x}>10^{43}$ erg s$^{-1}$) appears to be restricted to a clearly defined 
strip that is different from the area where most of the 
star-forming galaxies are found.
This is shown in panel (b) of  Fig. \ref{fig:all_X}, where the two dashed lines indicate the 
particular strip of the SFG region where most of the X-ray objects are found. 
In the redshift range covered by the \blue\ diagram, there are no sources at 
L$_{\rm X}$$<10^{42}$ erg s$^{-1}$ due to the flux limit of the X-ray observations.

\begin{figure}[h!]						     
\begin{center}                                                      
\includegraphics[height=7.5cm,width=8.0cm]{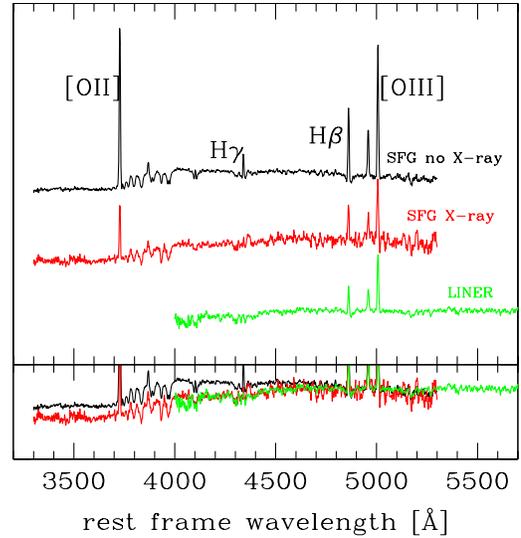}
\caption{Composite spectrum of X-ray sources optically classified as star-forming galaxies (red line) 
compared to the composite obtained with all the sources that are not X-ray emitting in the same strip of the star-forming region 
(black line) and to the composite obtained from the LINER sample selected at low redshift with the BPT diagrams (green line).
The lower panel shows the three spectra normalized in the wavelength range around the \OIIIb\ line.
The comparison between the red and the black line highlights different emission-line strengths (upper panel) 
and a redder continuum of the X-ray emitters (lower panel), suggesting an older stellar population component as well as  
possible dust extinction on galactic scales. In contrast, the X-ray emitting SFG show properties very similar to 
the LINER composite (red and green line) in terms of line intensities and continuum shape.}
\label{fig:composite}                                                     
\end{center}                                 
\end{figure}

Figure \ref{fig:composite} (upper panel) shows the composite spectra of SFGs lying in the 
``strip'' with detected X-ray emission (red line) and without X-ray signature (black line).
While the line ratios of the two composite spectra are indeed very similar, hence their location in the 
same region of the  diagnostic diagram, important differences should be noted.
Firstly, the X-ray sources have far weaker emission lines than the non X-ray sources (top panel).
Secondly, the normalized representation in the lower panel indicates that the X-ray sources have 
a significantly redder continuum which can be interpreted as an older stellar population in the host 
galaxy and/or as possible dust extinction on galactic scales. 

%The composite spectra of the SFGs lying in the ``strip'' with (red line) 
%and without (black line) detected X-ray emission are presented in Figure \ref{fig:composite} (upper panel).\\
%The most evident difference between the two composite spectra is that, even having very similar 
%lines ratios which indeed locate them in the same region of the diagram, the X-ray sources have 
%much weaker emission lines compared to the non X-ray ones.  
%Moreover, as shown in the lower panel of the same figure, the composite of X-ray sources shows a redder 
%continuum with respect to the non X-ray sources, \textbf{which can be interpreted as a sign of an 
%older stellar population in the host galaxy as well as possible extinction on galactic scales. 
However, the composite spectrum of the SFGs with X-ray emission has very similar 
properties, in the common spectral range, to the composite obtained from the sample of LINERs 
(green line) selected at low redshift using the 
BPT diagrams. As shown in Fig. 
\ref{fig:composite}, they have very similar lines intensities (upper panel) and continuum shape 
(lower panel).  
Given these similarities, our interpretation is that many of these X-ray emitting sources in the SFG region could be  
misclassified LINERs. This is unsurprising given the selection within the \blue\ diagram, which
corresponds to a nearly flat cut in \OIIIb/\Hb, given the range of 
\OII/\Hb\ probed by our sample. Applying a similar flat cut 
(\OIIIb/\Hb $>$6) to the low-z sample, 
we would have failed to identify LINERs \citep[see also][]{Lamareille2009dd}.

However, there is a second hypothesis that we should consider. These X-ray sources in the SFG region could 
also be composite AGN/SF objects in which star formation and AGN activity coexist, as expected in the 
current framework of galaxy-AGN co-evolution models. 
This second hypothesis is consistent with model predictions of the source position 
in the optical diagnostic diagrams.
%From these evidences, we can conclude that either
%\begin{itemize}
%\item[(a)] these sources are misclassified by the \blue\ diagnostic diagram and are thus normal AGN, which
% would mean that this diagnostic is not efficient in selecting AGN and that there is a particular 
%area in the SFG region where AGN can be found; or
%\item[(b)] these sources are AGN extinct on galactic scales, as suggested by the redness of the
%continuum in the optical composite spectrum (see Figure \ref{fig:composite}) and considering that the \blue\ diagram is 
%known to be sensitive to reddening effects they are not recognized as AGN; or finally  
%\item[(c)] these sources are composite objects in which star formation and AGN activity coexist as
%expected in the current framework of galaxy-AGN coevolution models.
%This hypothesis does not contradict with their position in the Fx/F$_{[O\,\textsc{iii}]}$ 
%vs F$_{[O\,\textsc{iii}]}$/F$_{IR}$ diagram (see figure 3):
%Panessa et al. (2005) indeed showed that composite objects also fall in the AGN region.
%\end{itemize}
\citet{Stasinska2006} %modeling the position of observational points 
%in the optical diagnostic diagrams, 
showed that while in the \reda\ diagram the separation line between AGN and SF is clearly 
defined in terms of the minimum AGN fraction,  
%the theoretical tracks in the \reda\ diagram of objects 
%with an increasing AGN fraction lie almost perpendicular to the division line (the right ``wing'' in the 
%upper right panel of Figure 4 in \citealt{Stasinska2006}) and the 
%separation line is thus well defined in terms of the minimum AGN fraction. \\
in the \blue\ diagram %on the contrary, %the theoretical track moves vertically away from the 
%locus of data points towards the sparsely populated center of the diagram.
galaxies with a moderate AGN fraction still lie in the star-forming locus. Hence, using this 
diagram objects will be  
observationally classified as AGN only when the AGN contribution is high. % (upper left panel of Figure 4 
%in \citealt{Stasinska2006}).}  
Based on these theoretical models, the existence of composite objects in the hashed region 
of the \blue\ diagram is plausible. 
%although the observed data points do not clearly outline the predicted model track.
%have suggested that objects that lie along the right wing 
%(upper right part) of the red1 diagram 
%differ mainly in the balance between massive stars and AGN ionizing powers. 
%They also showed that the same sequence in the blue diagnostic diagram falls towards the left.
%They thus demonstrated that composite SF-AGN objects lie 
%in the left part of the SF region of the blue diagram until the AGN contribution is relatively high. 
%Although the position in our diagram does not perfectly match the track found in \citet{Stasinska2006} 
%the hypothesis of composite objects is well established in this framework.
%_______________________________________________________________
%_______________________________________________________________

New IR spectroscopic observations have been obtained with SofI, the infrared 
spectrograph and imaging camera on the NTT, for a larger sample of objects with the same 
properties based also on COSMOS-Chandra data (\citealt{Elvis2009}, Civano et al., in prep).  
The IR spectra combined with the multi-wavelength information 
available in COSMOS will allow us to ascertain more accurately the true 
nature of these objects. A more detailed discussion of these data will be presented in a forthcoming paper.

For the purposes of this paper, we decided not to include these sources 
(i.e. X-ray detected, but optically classified SFG) in our AGN sample.  
Given the observationally well known differences between the X-ray  
properties and the optical spectral types \citep{Trouille2009} a mixed classification scheme can 
complicate the interpretation.

\section{Luminosity function} \label{sec:lf_method}

To study the evolution of type--2 AGN, we derived the luminosity function, 
which describes the number of AGN per unit volume and unit luminosity in our sample.
Since the optical continuum of type--2 AGN is dominated by the host galaxy, to sample 
and study only the AGN we have to rely on the luminosity derived from the 
emission lines connected to the ionizing source (the AGN in the core). 
We decided to use the \OIIIb\ emission-line because (1) the contamination from star 
formation is small 
for this line and thus its luminosity reflects the true AGN contribution more accurately 
than any other line \citep{Hao2005lf}, and 
(2) the \OIIIb\ line is by construction present in all our spectra

%after adding the brighter SDSS sample, spanning the same redshift range, 
%we derived a model to describe the evolution in redshift of this class of objects in different
%luminosity range.
 
%\subsection{\lOIIIb\ line luminosity}

We derived the \OIIIb\ luminosities from the emission-line fluxes measured by 
the automatic pipeline platefit\_vimos (\citealt{Lamareille2009}; Lamareille et al., in prep).
%as:
%\begin{equation}
%\rm L_{[OIII]}=\rm f_{[OIII\lambda5007\AA]} \ 4 \pi \times \rm D_{\rm L}(\rm cm)^2 
%\end{equation}
%where D$_{\rm L}$(cm) is the luminosity distance of the object expressed in cm and \rm f$_{\rm \OIIIb}$ is the \OIIIb\ 
%flux, as measured by the automatic pipeline platefit\_vimos (\citealt{Lamareille2009}; Lamareille et al., in prep).
We did not correct the \OIIIb\ flux for aperture effects.   
This correction would take into account the fraction of light 
of a given source that was missed because of the finite width of the slits in the VIMOS 
masks (1 arcsec). This factor is close to 1 (corresponding to no correction) for stars and 
increases towards more extended objects. For our sample of host galaxies of type--2 AGN, the correction 
factor for the continuum 
ranges between 1 and 3 with an average value of 2.2. 
%However, since we wish to estimate the light 
%coming from the \textbf{AGN narrow-line region and avoid contamination by star formation occurring 
%in the disk}, we decided not to apply this correction.
However, if our AGN classification for these objects is correct, most of the \OIIIb\ luminosity 
is produced in the AGN narrow-line region (which has a characteristic scale of $\rm 2-10\ kpc$; \citealt{Bennert2002}) 
and should therefore be treated as a compact source. 
For this reason, we did not apply any slit-loss correction to the observed \OIIIb\ fluxes.

%corrected for aperture effects. This correction takes into account the fraction of light 
%of a given source that has been missed because of the width of the slits in the VIMOS 
%masks (1 arcsec).
%The aperture correction is computed from Subaru and ACS i-band with the following procedure:
%the spectrum is convoluted with the considered filter (ACS and subaru i-band) and the 
%derived magnitude is compared with the I-band magnitude derived from the images. 
%The difference between the two magnitudes is the aperture correction factor. 
%The correction factor spans from xx to xx and for most of the objects (xx\%) is xx.

\subsection{Incompleteness function} \label{sec:completeness}

To study the statistical properties of type--2 AGN, we first need to derive the total number 
of type--2 AGN in the field and we therefore need to correct our sample for the fraction of objects 
that are not included because of selection effects.
In particular, we correct for the sources that were not observed spectroscopically 
(\textit{target sampling rate}, hereafter TSR) and for those that were not identified from 
their spectra (\textit{spectroscopic success rate}, hereafter SSR).
In particular, the TSR is the fraction of sources observed in the spectroscopic
survey compared to the total number of objects in the
parent photometric catalogue. 
As a general strategy, sources are selected randomly without any bias. However, some particular 
object (e.g., X-ray and radio sources) are designated compulsory targets, i.e., objects upon which 
slit must be positioned. The TSR in the latter case is much higher ($\sim$ 87\%) than for the random sample ($\sim$ 36\%). 
The SSR is the fraction of spectroscopically targeted objects 
for which a secure spectroscopic identification was obtained. It is computed to be the ratio 
of the number of objects with measured redshifts to the total number of spectra 
and ranges from 97.5\% to 82\% as a function of apparent magnitude. 
Therefore, the incompleteness function consists of two terms linked to (a) the
selection algorithm used to design the masks and (b) the quality of the spectra, respectively. 
The correction is performed using a statistical weight associated with 
each galaxy that has a secure redshift measurement. This weight is the product of the inverse of 
the TSR (w$^{\rm TSR}$=1/TSR) and of the SSR (w$^{\rm SSR}$=1/SSR) and was derived by 
\citet[][see also \citealt{Bolzonella2009}]{Zucca2009} 
for all objects with secure spectroscopic redshifts, taking into 
account the compulsory objects\footnote{In the selected type--2 AGN sample, 17 sources were compulsory.}.

%_____________________________________FIGURE_______________________________
\begin{figure}                                                      
\begin{center}                                                      
\includegraphics[height=8.0cm,width=8.0cm]{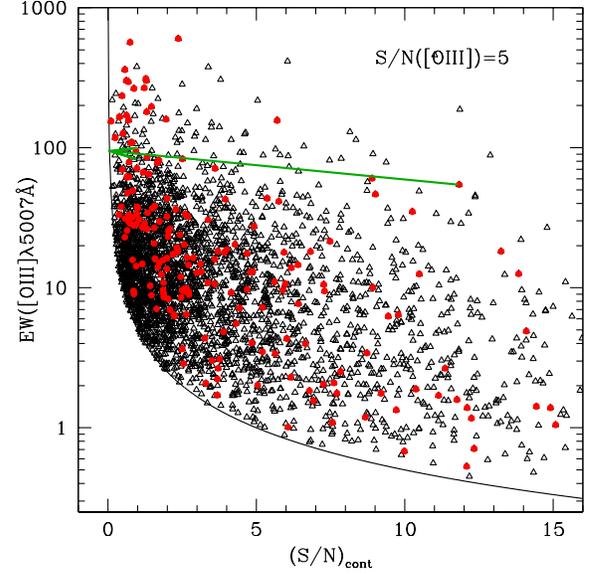} 
\caption{Observed EWs of the \OIIIb\ emission line
versus the signal-to-noise ratio of the continuum close to the line.
Black triangles correspond to the emission-line sample with measured \OIIIb\ S/N$>$5, while red 
circles highlight the type--2 AGN sample. 
The lower envelope represents the cut in S/N of the line (S/N$_{\rm \OIIIb}>$5) and 
gives the minimum EW detectable given the S/N of the continuum.
The green arrow traces, as an example, the position of a given object in this plane for increasing redshift. The
z$_{max}(f_{i,l})$, used to compute the V$_{max}(f_{i,l})$, corresponds to the point at which the source reaches the solid line.}
\label{fig:ew_sn_vmax}                                                     
\end{center}                                 
\end{figure}
%____________________________________________________________________________________

%_____________________________________TABLE______________________________

\begin{table}
\begin{center}
\begin{tabular}{c c c c c c }
\hline
\hline
\multicolumn{2}{c}{$\Delta\log$L$_{\rm \OIIIb}$[L$_\odot$]}&N$_{\mathrm{AGN}}$&$\log\Phi$(\OIIIb)&\multicolumn{2}{c}{$\Delta\log\Phi$(\OIIIb)}\\
\hline
\multicolumn{6}{c}{$0.15 < z < 0.3$}\\
\hline
\noalign{\smallskip}
  5.60  & 6.10  &  10  & -3.42  &  +0.12 &   -0.17\\
  6.10  & 6.60  &  11  & -3.40  &  +0.12 &   -0.16\\
  6.60  & 7.10  &   3  & -4.08 &   +0.20  &  -0.34\\
  7.10  & 7.60  &   1  & -4.49  &  +0.52  &  -0.76\\
  7.60  & 8.10  &   1  & -4.95 &   +0.52 &   -0.76\\
\hline
\multicolumn{6}{c}{$0.30 < z < 0.45$}\\
\hline
  5.43 &  5.93  &  5	& -3.87  &  +0.20  &  -0.37\\
  5.93 &  6.43  & 38	& -3.17 &   +0.07  &  -0.09\\
  6.43 &  6.93  & 34	& -3.27 &   +0.07  &  -0.09\\
  6.93 &  7.43  & 17	& -3.66  &  +0.10  &  -0.12\\
  7.43 &  7.93  &  6	& -4.23  &  +0.16  &  -0.25\\
  7.93 &  8.43  &  1	& -4.92  &  +0.52  &  -0.76\\
  8.43 &  8.93  &  1	& -4.92  &  +0.52  &  -0.76\\
\hline
\multicolumn{6}{c}{$0.5 < z < 0.92$}\\
\hline
  6.32 &  6.82 &   3   & -4.54 &  0.20  &  0.40 \\
  6.82 &  7.32 &   15  & -4.04 &  0.11  &  0.16 \\
  7.32 &  7.82 &   36  & -3.83 &  0.08  &  0.10 \\
  7.82 &  8.32 &   20  & -4.18 &  0.11  &  0.15 \\
  8.32 &  8.82 &    7  & -4.77 &  0.15  &  0.23 \\
  8.82 &  9.32 &    4  & -5.16 &  0.18  &  0.32 \\
\hline
\hline
\end{tabular}
\end{center}
\caption{Binned $\log\Phi$(\OIIIb) luminosity function estimates  for $\Omega_m$=0.3, $\Omega_{\Lambda}$=0.7, and 
H$_0$=70 km $\cdot$ s$^{-1}$ $\cdot$ Mpc$^{-1}$.
We list the values of $\log\Phi$ and the corresponding 1$\sigma$ errors in three redshift ranges, 
as plotted with full circles in Fig. \ref{fig:lf} and in $\Delta \rm log(L_{\rm \OIIIb}[L_{\odot}])$=0.5
luminosity bins. We also list the number of AGN contributing to the luminosity function estimate 
in each bin.}
\label{tab:lf}
\end{table}

%____________________________________________________________________________________

\subsection{1/V$_{max}$ luminosity function}

We derive the binned representation of the luminosity function using the usual
$1/V_{\rm max}$ estimator \citep{Schmidt1968}, which gives the space-density
contribution of individual objects. 
The 1/V$_{\rm max}$ method considers for each object $i$ the comoving 
volume (V$_{\rm max}$) within which the $i^{th}$ object would still be included in the sample.
To calculate $V_{\rm max}$, we thus need to consider how each object has been selected to be in our 
final sample of 213 sources.

The zCOSMOS type--2 AGN sample was derived from a magnitude-limited sample 
after applying a cut to the S/N ratio of the appropriate emission lines in the 
diagnostic diagram adopted. The V$_{\rm max}$ of each object is thus linked to the maximum 
apparent magnitude as well as the minimum flux of the involved lines.  

While the maximum magnitude is the same for all the objects (by definition for 
the zCOSMOS bright sample I$_{\rm AB}<$22.5), the definition of the minimum flux of the lines 
differs for each object depending on the continuum level.

Following the procedure described by \citet{Mignoli2009}, for each object we estimate the emission-line 
detection limit considering the S/N in the continuum adjacent to the line and assuming the cut applied to 
the S/N of the line. 
Figure \ref{fig:ew_sn_vmax} shows, as an example, the result of this procedure for the \lOIIIb\ line. 
In this plot, we show the observed \OIIIb\ EW 
as a function of the continuum S/N for both the entire emission-line sample (black triangles) 
and the type--2 AGN sample (red circles).
The solid line represents the cut to the S/N of the \OIIIb\ line (S/N$_{\rm \OIIIb}>$5) and 
indicates the minimum EW detectable given the S/N of the continuum.
In this plane, sources move diagonally (left and upwards) towards the solid line going to higher 
redshift, since the observed EW of the line increases with redshift as the continuum signal decreases. 
The green arrow in Fig. \ref{fig:ew_sn_vmax} 
traces, as an example, the evolution with redshift of the position of a given object in this plane.
At a given redshift z=z$_{max}$, the source reaches the minimum S/N detectable and thus the same object 
at z$>$z$_{max}$ would not have been included in our sample because of the cut applied to the S/N of the \OIIIb\ line. 
This procedure allows us to compute for each object the V$_{max}$ relative to a given line as the 
volume enclosed between z=0 (or z=z$_{min}$) and the derived z$_{max}$.
 
The same procedure was repeated for all the emission lines $l$ used in the selection (the line S/N cut is 5 for 
\OIIIb\ but 2.5 for all the other lines) resulting, for each object, in a number of V$_{max}(f_{i,l})$, each 
corresponding to a different line.

Finally, the maximum volume for each object $i$ was estimated to be the minimum between the volume 
$V_{max}(m_i)$ associated with the maximum apparent magnitude and the volumes $V_{max}(f_{i,l})$ associated 
with the minimum flux of the used lines.

\begin{figure*}                                                      
\begin{center}                                                      
\includegraphics[height=16.0cm,width=16.0cm]{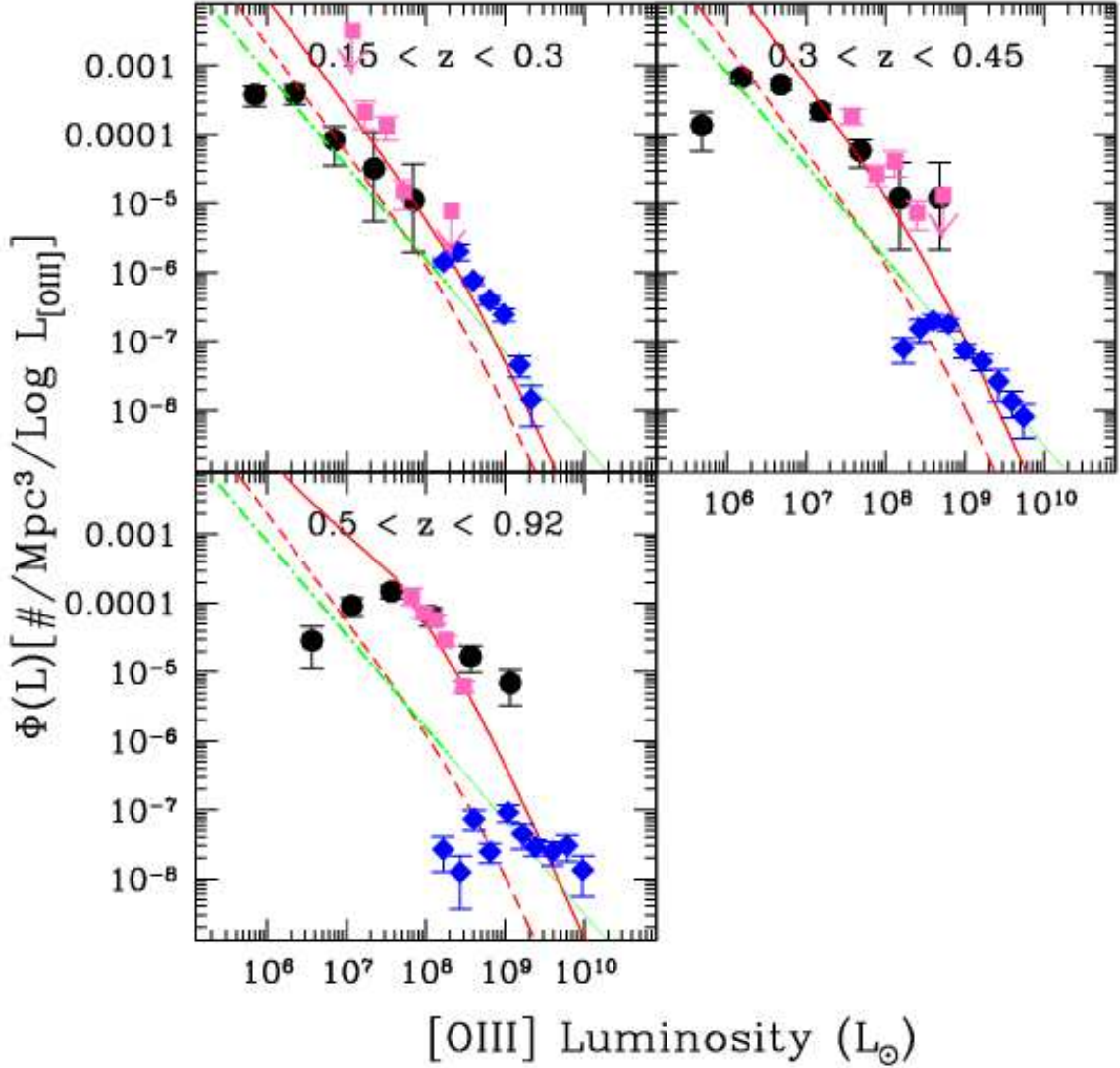} 
\caption{Binned \OIIIb\ line luminosity function of the zCOSMOS type-2 AGN (black circles) derived in the redshift bins
0.15$<z<$0.3, 0.3$<z<$0.45, and  0.45$<z<$0.92, compared to the SDSS (R08) type--2 (blue diamonds)  AGN data.
Pink squares show the [2-10keV] LF derived for the entire (obscured and unobscured) XMM-COSMOS AGN sample 
and converted to \OIIIb\ luminosities using the \citet{Netzer2006} relation.
The curves in the figure show LF models derived by other authors.
Dot-dashed and dashed lines show the local (z=0) LF derived from an optically 
selected sample \citep[green dot-dashed line;][]{Hao2005lf}  
and from an X-ray selected sample (red dashed line; DC08), respectively. 
Moreover, in each panel the LF model from 
DC08 evolved to the mean redshift of the bin is reported with a solid red line. 
The X-ray LF from DC08 was converted to a \OIIIb\ LF 
using the mean \OIIIb\ to X-ray luminosity ratio 
derived by \citet{Mulchaey1994} (L$_{\rm \OIIIb}$[L$_{\rm \odot}$] $\simeq$ 3.907 
$\times$ 10$^{6}$ L$_{\rm x}$[10$^{42}$]).
%The green line in the figure shows the local (z=0) LF model derived from an optically 
%selected sample \citep{Hao2005lf}. 
%To convert from X-ray [2-10]keV to \OIIIb\ luminosities we used the mean ratio 
%derived by \citet{Mulchaey1994} (L$_{\rm \OIIIb}$[L$_{\rm \odot}$] $\simeq$ 3.907 $\times$ 10$^{6}$ L$_{\rm x}$[10$^{42}$]).
}
\label{fig:lf}                                                     
\end{center}                                 
\end{figure*}

The luminosity function for each redshift bin ($z-\Delta z/2$ ; $z+\Delta z/2$) is thus computed to be
\begin{equation}
\Phi(L) = \frac{1}{\rm \Delta \log L} \sum_i \frac{\rm w^{TSR}_i 
w^{SSR}_i}{\min[\rm V_{max}(m_i),V_{max}(f_{i,l})]},
\end{equation}
where $w^{TSR}_i$ and $w^{SSR}_i$ are the statistical weights described above.\\ 
The statistical uncertainty in $\Phi$(L) is given by \cite{Marshall1983a}

\begin{equation}
\sigma_{\phi} = \frac{1}{\rm \Delta \log L} \sqrt{\sum_i \frac{\rm (w^{TSR}_i 
w^{SSR}_i)^2}{\min[\rm V_{max}(m_i),V_{max}(f_{i,l})]^2}}
\end{equation}
The resulting luminosity functions in different redshift ranges are shown in
Fig. \ref{fig:lf}, while the details for each bin are presented in Table 
\ref{tab:lf}. 
%These values are obtained assuming a standard cosmology with 
%$\Omega_{\rm m}$ = 0.3, $\Omega_{\Lambda}$ = 0.7 and H$_{0}$ = 70 km s$^{-1}$
%Mpc$^{-1}$. 

%____________________________FIGURA_________________________________________

%-----------------------------
%____________________________FIGURA_________________________________________

\begin{figure*}                                                      
\begin{center}                                                      
\includegraphics[height=16.0cm,width=16.0cm]{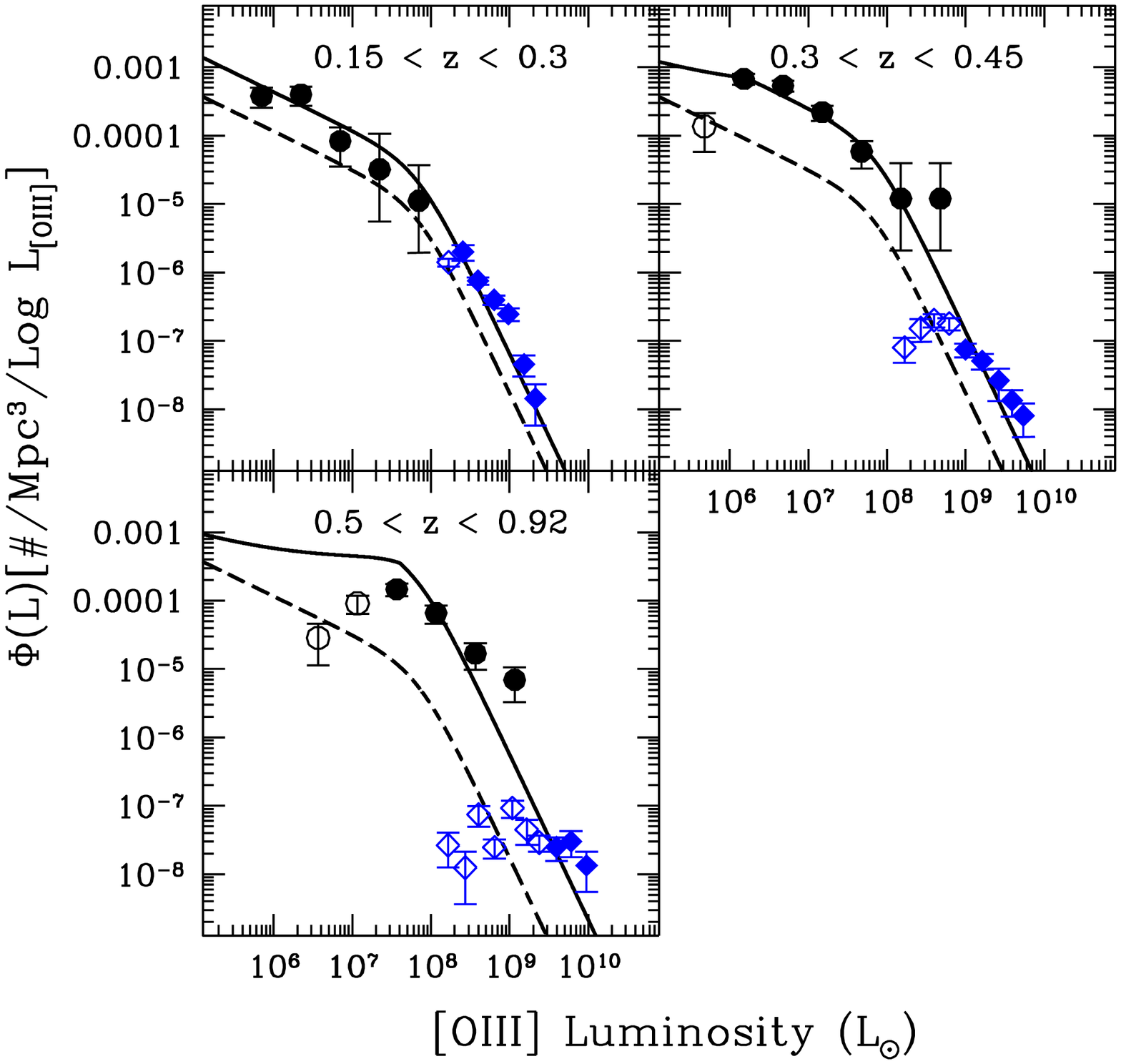} 
\caption{Binned \OIIIb\ line luminosity function of zCOSMOS (black circles) and SDSS 
(blue diamonds) type-2 AGN samples. Open symbols show incomplete bins (see Sect. \ref{sec:lf_result}) 
which were not used to derive the model fits.
The black curve in the figure shows the LF best fit model (LDDE) derived 
considering the combined zCOSMOS-SDSS data shown here. 
In each panel, the z=0 model is also reported as reference (dashed line).}
\label{fig:LF_model}                                                     
\end{center}                                 
\end{figure*}
%-----------------------------

\section{Results} \label{sec:lf_result}

Figure \ref{fig:lf} shows our LF data points (black circles) and, for comparison, the binned luminosity function 
derived from the SDSS sample of type--2 AGN (blue diamonds) in the same redshift 
range from R08. 
The last redshift bin of this figure corresponds to the redshift range spanned by the \blue\ 
diagnostic diagram that, as discussed in Sect. \ref{sec:Xcomp}, may be affected by a  
more significant incompleteness, which is considered below.

As already pointed out, the SDSS sample is complementary in terms of \OIIIb\ luminosity to the 
zCOSMOS sample and spans a similar redshift range. Thus, combining the two samples 
allows us to constrain the luminosity function over a wide luminosity range.
As can be seen in Fig. \ref{fig:lf}, for at least the first two bins the two data sets are
in good agreement, with the zCOSMOS points connecting smoothly to the bright SDSS data points.

In Fig. \ref{fig:lf}, it is also shown (pink squares) as comparison the X-ray luminosity function data
(Miyaji et al., in prep) derived in the same field using the XMM-COSMOS sources \citep{Cappelluti2009} with optical 
identifications by Brusa et al. (in prep).
The XMM-COSMOS LF data points were converted from X-ray [2-10]keV to \OIIIb\ luminosities 
by applying the luminosity dependent relation (log(L$_{\rm \OIIIb}$/L$_{2-10})$ = 16.5 - 0.42 logL$_{2-10}$)
derived by \citet{Netzer2006}. 

The XMM-COSMOS LF overlaps with our luminosity range and is in very good agreement with our LF data points 
showing, in some of the bins, a higher density.   
This is not surprising since the X-ray LF refers to the entire AGN population (obscured and unobscured), 
while our LF considers only obscured sources.
However, a one-to-one correspondence between X-ray and optical classification does not hold since these bands select 
different populations, with e.g., X-ray surveys missing Compton thick AGN \citep{LaMassa2009}.  

%However, as we pointed out several times,   
%and populates also the intermediate luminosities 
%where a gap between zCOSMOS and SDSS data is present. 

%\textbf{The X-ray LF, referring 
%to the entire AGN population (obscured and unobscured) is higher, especially in the frfirstist two redshift bins, 
%than our LF data points of only obscured sources.}
%especially in the first two bins,  
%Despite it refers to the entire AGN population (obscured 
%and unobscured), it is in good agreement with our LF data points 
%suggesting that at these luminosities and redshifts type--2 AGN are a significant fraction 
%of an X-ray selected sample of AGN.   

%Figure \ref{fig:LF_model} together with the zCOSMOS (black circles) and the SDSS (blue diamonds) data.
For all redshift bins, the faintest part (first data points) of our LF 
evidently declines.
This is 
%probably 
an artifact related to the selection of the zCOSMOS sample, which 
is based on broad-band magnitude (I$_{\rm AB}<$22.5). 
%statistically complete for all objects with I$_{AB}<$22.5.
%This selection based on broad-band magnitude only is not well correlated with the strength of 
%the \OIIIb\ emission, used for our LF. 
This implies that a fraction of objects that would fulfill our \OIIIb\ based cuts, 
%might be 
are too faint in the I$_{\rm AB}$ band to be included in the zCOSMOS sample.
These objects never enter the sample, even at the minimum redshift, so 
they cannot be corrected.
% missed fraction of objects cannot be quantified with the available data set and is thus not 
%accounted for in our completeness estimate in Section \ref{sec:completeness}.
 % could be due, at least partly, to the fact that while the LF has been computed 
%in \OIIIb\ line luminosity and thus assuming a \OIIIb\ line flux limited sample, 
%the sample was originally a total I$_{AB}$ apparent magnitude selected sample I$_{AB}$<22.5. 
%This lead to an incompleteness that is not taken into account: there are in fact objects 
%that should be inside the \OIIIb\ line flux limited sample because 
%they are above the \OIIIb\ flux threshold but they are missed because their 
%apparent total magnitude was too faint to be originally included in the zCOSMOS 10k sample. 
%With the plausible assumption that 
Since intrinsically faint objects tend to be fainter in \OIIIb, 
the fraction of missed type--2 AGN is thus higher in the lowest \OIIIb\ luminosity bins.
%and increases with redshift because of cosmological dimming effects. 
The onset of significant incompleteness can be approximately estimated by the following back-of-the-envelope calculation. 
We convert from the limiting apparent I-band magnitude (I$_{\rm AB}<$ 22.5) to an absolute magnitude at 
the upper bound of the redshift bin. Using the median EW in the redshift bin, we then estimate the absolute \OIIIb\ luminosity at which 
about half of the objects at this redshift should be missing. Applying this to our data set, we found good consistency 
between the estimated onset of incompleteness and the position at which the LF begins to turn over. 
In particular, we found that the approximate \OIIIb\ luminosities where significant incompleteness 
is expected are $\sim$ 9$\times$10$^5$ L$_{\odot}$,  $\sim$ 2$\times$10$^6$ L$_{\odot}$, and $\sim$ 10$^8$ L$_{\odot}$ 
for z = 0.3, 0.45, and 0.92, respectively (first, second, and third redshift bin). The incomplete bins will 
not be taken into account in the computation of the model to describe the \OIIIb\ LF and 
its evolution in Sect. \ref{sec:lf_model}.

On the other hand, the possible absence in our AGN sample of misclassified LINERs in the \blue\ diagram 
 would affect the LF only in the last redshift bins.  
If most of the 31 X-ray detected sources located in the region of star-forming galaxies in the \blue\ 
diagram are indeed AGN (see Sect. \ref{sec:Xcomp}), assuming that the fraction of X-ray detections for 
them is similar to that of Sey--2 galaxies (from 10\% to 20\%; see Table \ref{tab:statistic}), 
we can estimate that the total number of misclassified objects 
in the star-forming region of this diagram is (5 - 10)$\times$ 31 $\sim$ 150 - 300 i.e., $\sim$ 5 - 10\% of 
all objects. Considering the distribution in \OIIIb\ luminosities of the X-ray detected 
sources possibly misclassified AGN  (L$_{\rm \OIIIb} \sim 3\times10^{6} - 5\times10^{7}$ L$_{\odot}$), 
and adding these sources to the corresponding affected luminosity bins,  
we find that the number density could increase by more than one order of magnitude in the 
first bin and about half in the second.
%Since their luminosities are mostly distributed in the \OIIIb\ range 
%$\sim 3\times10^{6} - 5\times10^{7}$ L$_{\sun}$, their absence would affect mainly the faint luminosity bins.} 
%In particular, once adding these sources to the affected luminosity bins 
%(L$_{\rm \OIIIb} \sim 3\times10^{6} - 3\times10^{7}$ L$_{\sun}$), 
%the number density in these bins could increase by more than one order of magnitude for the first bin and 
%about half for the second.

Finally, we note that extinction could affect the whole LF shape and/or normalization 
shifting the data points towards fainter luminosities. 
The \OIIIb\ line is expected to be affected by dust extinction, 
located either within the narrow-line region itself or in the intervening interstellar 
matter of the host galaxy.
%Derive the extinction presents however many complications (as shown by previous studies of Balmer line 
%intensities; see \citealt{Anderson1970,Adams1975,Baldwin1975,Osterbrock1976,Osterbrock1977}).
Since the quality of the zCOSMOS spectra does not always allow a reliable estimate of extinction on an object-by-object basis, 
we decided not to apply any dust extinction correction to our \OIIIb\ luminosities.
However, we found that the H$\alpha$/H$\beta$ ratios measured on the composite spectra of Sey--2 
and LINERs are $\sim$2.55 and $\sim$2.45 respectively, consistent inside the errors with no extinction.
%The distribution of the H$\alpha$/H$\beta$ ratio has a large dispersion and is on 
%average consistent with no extinction with a bunch of objects 
%at L$_{\rm \OIIIb}< 5\times10^6$ with \Ha/\Hb$\sim$4.25, corresponding to an extinction of 1.2 magn 
%(for a Milky way extinction curve with R = 3.1).}
% To determine the extinction we should consider the Balmer 
%line ratios. However to do that, we would need high S/N for the lines H$\alpha$, H$\beta$, 
%H$\gamma$ and H$\delta$. Since the zCOSMOS spectra don't have the accuracy to perform 
%this kind of studies and since previous studies of Balmer line intensities have showed that 
%there are many complications in this derivation (they found that the Balmer lines 
%cannot be explained just considering a standard recombination theory and a standard extinction 
%law, \citet{Anderson1970,Adams1975,Baldwin1975,Osterbrock1976,Osterbrock1977}), we decided to not 
%apply any dust extinction correction to our \OIIIb\ luminosities.\\
%Since the correction would increase the \OIIIb\ luminosity, 
%using the uncorrected line luminosities our objects would have been assigned to 
%fainter luminosity bins and 
%the corrected LF would be shifted towards brighter luminosities.

The green line in all of our bins corresponds to the LF derived from the local SDSS sample 
\citep[$z<0.15$;][]{Hao2005lf}. As can be seen, in the first redshift bin the zCOSMOS and 
the SDSS data points are in good agreement with the fit to the local LF model suggesting that no detectable  
evolution occurs between z$\sim$0 and z$\sim$0.22. 

In contrast, in the second redshift bin, an evolutionary trend is clear and the combined 
zCOSMOS-SDSS data points seem to follow the same evolutionary model found by \citet[][ hereafter DC08]{DellaCeca2008} 
using an X-ray selected sample of obscured AGN from the XMM-Newton hard bright serendipitous 
sample (HBSS) with spectroscopic identification.
In this work, DC08 attempted to fit   
a luminosity-dependent density evolution (LDDE) model similar to and consistent with previous work 
\citep{Hasinger2005,LaFranca2005}.
%We converted their LF assuming the mean 
%L$_{\rm \OIIIb}$/L$_{(2-10)keV}\simeq$ 0.015 ratio for Seyfert galaxies obtained by \citet{Mulchaey1994}, 
%fully consistent with the value reported in \citealt{Heckman2005} for the unobscured view of Seyfert 
%galaxies, L$_{\rm \OIIIb}$/L$_{(2-10)keV}\simeq$ 0.017. 
We overplot
their local (z=0) and evolved LF appropriately transformed into our figure as dashed and solid 
red lines, respectively. As shown in the figure, the zCOSMOS-SDSS data points in the second redshift bin, 
lie along the solid line and indeed follows a similar trend.
For this curve, the conversion from X-ray to \OIIIb\ luminosities
was performed by assuming the mean 
L$_{\rm \OIIIb}$/L$_{(2-10)keV}\simeq$ 0.015 ratio for Seyfert galaxies obtained by \citet{Mulchaey1994} 
(fully consistent with the value reported in \citealt{Heckman2005} for the unobscured view of Seyfert 
galaxies, L$_{\rm \OIIIb}$/L$_{(2-10)keV}\simeq$ 0.017).
The luminosity dependency of the \citet{Netzer2006} relation would cause a discrepancy with our data points, 
especially at the bright end. However, since this relation was derived for a more limited \OIIIb\ 
luminosity range, its application to our objects with the highest \OIIIb\ luminosity would correspond 
to an extrapolation of the relation beyond the original data range.

At higher redshift (z$\sim$0.7), the agreement is no longer as good as in the other two bins, but 
(see Sect. \ref{sec:Xcomp}) in this redshift range the optical and the X-ray selections 
do not sample the same population and a direct comparison between them is thus not possible. 
The SDSS LF data points in this redshift bin also show a significant incompleteness:  
R08 highlighted that because of different selection biases,  
their highest quality data at high redshift (0.50$<z<$0.83) correspond to high luminosities 
(L$_{\rm \OIIIb} >$10$^{9.5}$ L$_{\odot}$). 
Our data points, compared to the X-ray model, show an excess of sources at high luminosities, 
while at low luminosities our data are probably affected by the incompleteness described above.
%due to e.g. the lack 
%of reddened AGN and/or composite SFG/AGN which are missed by the optical selection (see Section \ref{sec:Xcomp}).    
However, our three central LF data points support the trend seen for the previous bin, 
showing an evolution consistent with the LDDE model from DC08.

\subsection{Model fitting} \label{sec:lf_model}

Given the wide luminosity range spanned by placing zCOSMOS and SDSS data together, we tried to derive 
a model to describe the \OIIIb\ LF and its evolution. 
%To obtain better constraints, especially at intermediate luminosities where a gap between 
%zCOSMOS and SDSS data is present, we also added the X-ray luminosity function derived in the same field using 
%the XMM-COSMOS sources \citep{Cappelluti2009} with optical identifications by \citet{Brusa2007}.
%The XMM-COSMOS data (see Brusa et al., in prep. and Miyaji et al., in prep for more details) 
%has been appropriately converted from X-ray [2-10]keV to \OIIIb\ luminosities and are overplotted as purple squares in 
%Figure \ref{fig:LF_model} together with the zCOSMOS (black circles) and the SDSS (blue diamonds) data. 

In the computation of the model fit, we did not consider the luminosity bins (in both SDSS and zCOSMOS sample) 
that are likely to be incomplete (see Sect. \ref{sec:lf_result}). They are shown as open symbols in Fig. \ref{fig:LF_model}.\\
To be sure that the model fit is not strongly influenced by the last redshift bin, which may be highly incomplete,
we computed the model fits presented below first by not including the data in this redshift bin, and 
then considering them, and we found that
the resulting parameters agree to within the statistical errors. The results reported below correspond to the entire 
redshift range (0.15$<z<$0.92).  

For all analyzed models, we parameterized the luminosity function as a
double power-law  given by
\begin{equation}
\Phi(L,z) = \frac{\Phi_L^*}{(L/L^*)^{\gamma1} + (L/L^*)^{\gamma2}},
\label{eq:phil}
\end{equation}
where $\Phi^*_L$ is the characteristic AGN density in Mpc$^{\rm -3}$, L$^*$  is the characteristic
luminosity, % around which the slope of the luminosity function is changing
and $\gamma1$ and $\gamma2$ are the two power-law indices. 

After attempting different model fits (i.e., pure luminosity evolution (PLE), pure density evolution (PDE), or a combination 
of luminosity and density evolution), 
we assumed a luminosity-dependent density evolution model (LDDE) with the parameterization introduced by 
\citet{Ueda2003}.  
%The LDDE model allows the redshift of the AGN density peak to change as a function of
%luminosity. 
From X-ray studies, it is now well established that a LDDE model provides a 
more accurate description of the evolutionary properties of AGN \citep{Hasinger2005,Ueda2003,
LaFranca2005,Silverman2008,Ebrero2009}, and this is the case also in the optical domain 
for at least type--1 AGN \citep[][ hereafter B07]{Bongiorno2007}.
%Moreover, our data showed a trend similar to the LDDE model 
%fit derived by DC08.\\
We can describe the LF as a function of redshift with 

\begin{equation}
\Phi(L,z)=\Phi(L,0)\cdot e(z,L),
\end{equation}
where

\begin{equation}
e(z,L) = \left\{ 
	\begin{array}{ll}
	(1+z)^{p1} & (z \leq z_{\rm c}) \\ 
        e(z_{\rm c},L)[(1+z)/(1+z_{\rm c})]^{p2} & (z > z_{\rm c})\\
	\end{array}
       \right. , 
\label{eq:ldde1}
\end{equation}
along with

\begin{equation}
z_{\rm c}(L) = \left\{ 
	\begin{array}{ll}
	z_{\rm c,0}& 
	(L \geq L_{a}) \\ 
        z_{\rm c,0}(L/L_{a})^{\alpha} & (L < L_{a})\\
	\end{array}
       \right. ,
\label{eq:ldde2}
\end{equation}
where z${\rm _c}$ corresponds to the redshift at which the evolution changes.
We note that in this representation z${\rm _c}$ is not constant but depends on luminosity.
This dependence allows different types of evolution to occur at different luminosities and can indeed reproduce
the differential AGN evolution as a function of luminosity, thus modifying the 
shape of the luminosity function as a function of redshift.  

Given the small redshift range covered by the data, we are unable to fully constrain the evolution. 
For this reason, we fixed the evolutionary parameters (p1, p2, $\alpha$, z$_{\rm c,0}$, L$_{a}$) 
and used the $\chi^2$ minimization method to derive the normalization $\Phi^*_L$, 
the characteristic luminosity L$^*$, and the bright and faint end slopes of the LF ($\gamma1$ and $\gamma2$).

The evolution parameters were fixed using the results obtained by DC08, appropriately 
converted to \OIIIb\ luminosity and our cosmology.  
By fixing p1=6.5, p2=-1.15, L$_{a}$=8.15$\times$10$^{43}$erg s$^{-1}$, z$_{\rm c,0}$=2.49 and $\alpha$=0.2, we obtained
the best-fit model parameters $\gamma1$=0.56, $\gamma2$=2.42, and L$^*$=2.7$\times$10$^{41}$erg s$^{-1}$ with the normalization 
$\Phi^*_L$=1.08$\times$10$^{-5}$ Mpc$^{\rm -3}$.
%If we fix also the normalization ($\Phi^*_L$), the resulting fit would be significantly worse.

%, which is very similar to what has been found for optically selected type--1 AGN (B07).
%\footnote{Assuming the parameters 
%from \citet{Bongiorno2007} we would obtain $\gamma1$=2.95, $\gamma2$=0.96 and L$^*$=2.59 10$^{42}$.}
%\new{However, it is worth noting that although both type--2 AGN and type--1 AGN sources are described by a LDDE model 
%the dependence of the evolution on the luminosity seems to be less extreme for type--2 AGN. 
%This can be interpreted as an indication of a variable ratio between obscured and unobscured AGN as a 
%function of luminosity and/or redshift as will be shown later.}
%The representation of this best model fit is shown as solid red line in figure \ref{fig:LF_model},  
%where all the data used in the derivation of the model are shown as filled symbols.
%The dashed lines in each panel represent the model fit at z=0. 

%Moreover, the green dotted line represent the LF model derived for the VVDS type--1 AGN 
%sample \citep{Bongiorno2007}, converted using the method described below.  \\
%This model reproduces well the shape of the luminosity function in the first redshift bin 
%but it underestimates the data in the second and the last redshift bin. 

%We thus tried to leave also the normalization $\Phi^*_L$ free to vary and we fixed only 
%the evolutionary parameter. The derived normalization is now $\Phi^*_L$=2.55$\times$10$^{-5}$,
%with best fit parameters $\gamma1$=0.5, $\gamma2$=2.13 and L$^*$=1.1$\times$10$^{41}$. 

The representation of this best-fit model is shown as a solid line in Fig. \ref{fig:LF_model}, where
%\textbf{together with the zCOSMOS and SDSS LF data points derived in four redshift ranges. 
%In the plot, 
all the data used in the derivation of the model are shown as filled symbols.
The dashed line in each panel represents the best-fit model at z=0. 

This model represents reasonably well the data points reproducing 
the shape of the LF in the first two redshift bins, with a slight underestimation of the bright-end SDSS data points.
The last bin is
%\textbf{two bins are} 
not well fitted. In this redshift bin, the LF data points %\textbf{these redshift bins}, 
show an excess in the bright part of the LF. A possible bias could in principle be due to 
a higher mean redshift of 
the bright objects with respect to the central redshift of the bin due to the increasing space 
density of AGN with redshift. However, we tend to exclude this possibility because the four objects 
in the most deviant data point at the bright end of the LF have a mean redshift of z$\sim$0.67 
and are hence very close to the central redshift of the bin. 
%we  the cannot be explained by the known 
%incompleteness at this redshift %\textbf{at these redshifts} 
%`and thus remains unclear at this moment. 
Upcoming larger samples (e.g., the 20k zCOSMOS sample) will provide superior data statistics  
and hence an improved constraining power.

\begin{figure*}                                                      
\begin{center}                                                      
\includegraphics[height=16.0cm,width=16.0cm]{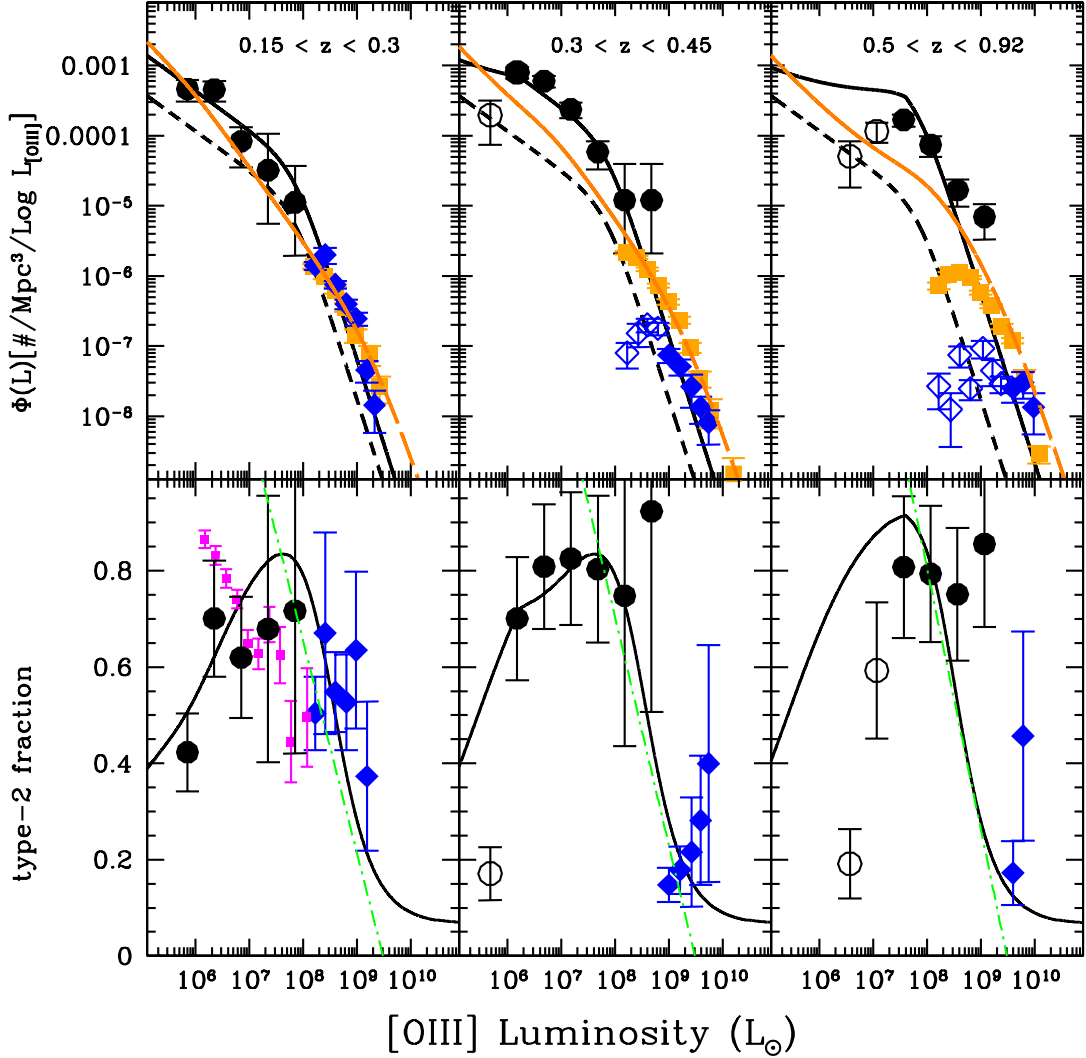} 
\caption{\textit{Upper panels:} Binned \OIIIb\ LF of type--2 AGN (black circles for zCOSMOS and 
blue diamonds for SDSS) and type--1 AGN (orange squares; R08); open symbols show incomplete bins not used to derive 
the model fit.
The curves represent the LF model fit: black for 
the type--2 AGN model as derived in Sect. \ref{sec:lf_model}; while orange for the VVDS type--1 AGN sample 
(B07) appropriately converted from broad-band luminosity to \OIIIb\ (see text).
\textit{Lower panels:} Fraction of type--2 AGN to the total type--1 + type--2 AGN population. 
Data points (black circles for zCOSMOS and blue diamonds for SDSS) 
are derived using the LF data points, while the black line is the resulting fraction considering the LF model fit for type--1 and type--2 AGN.
As a comparison, in the first redshift bin, the fraction of obscured AGN derived by \citet{Simpson2005} 
is overplotted with magenta squares. 
%The dotted red line shows the fraction of Compton-thin AGN assumed in 
%the \citet{Gilli2007} population synthesis model, while 
Finally, the dot-dashed green line is the linear fit to the type--2 fraction found by \citet{Hasinger2008} 
and converted using the \citet{Netzer2006} relation.}
\label{fig:fraction}                                                     
\end{center}                                 
\end{figure*}
%-----------------------------

\subsection{Type--2 AGN fraction} \label{sec:fraction}

One of the most important open issues regarding absorbed AGN is understanding their 
relevance amongst the AGN population and if there is a dependence of the fraction 
of absorbed AGN on either luminosity and/or redshift.

We computed the type--2 AGN fraction, i.e., the ratio of type--2 to total 
(type--1 + type--2) AGN, using the derived number densities for the type--2 AGN sample. 
In the analyzed redshift range (0.15$<z<$0.92), direct \OIIIb\ LF measurements for type--1 AGN are 
available only at $10^{8.3}$ L$_{\odot} <$ L$_{\rm [OIII]} <$ 10$^{10}$ L$_{\odot}$ (SDSS; R08). The zCOSMOS 
type--2 AGN luminosity regime remains thus mostly uncovered by data.  
To constrain also this luminosity range, we have converted the optical broad-band 
M$_{\rm B}$ LF derived by B07 
for type--1 AGN to a \OIIIb\ LF.  
%\textbf{Since for type--1 AGN the only \OIIIb\ LF data available are from the SDSS and 
%thus the zCOSMOS type--2 AGN regime} is not constrained by the data,  
%following the same approach used by R08 we converted the optical M$_{\rm B}$ LF derived by B07 
%for type--1 AGN to the \OIIIb\ LF. 
The B-band LF computed by B07 ranges from M$_{\rm B }= -20$ to M$_{\rm B} = -26$ thus probing 
the \OIIIb\ luminosity interval from $\sim$ 10$^{7.2}$ L$_{\odot}$ to $\sim$ 10$^{9.5}$ L$_{\odot}$ 
(see Eq. \ref{eq:MBLO3}).

To do this, following the same approach used by R08, we used the mean L$_{\rm\OIIIb}$-M$_{\rm B}$ relation 
and its scatter $\sigma$.
The \OIIIb\ luminosity is not a perfect tracer of bolometric luminosity and there is indeed 
substantial scatter between \OIIIb\ and continuum luminosity for type 1
quasars (\citealt{Netzer2006}, R08).
We thus considered the scatter $\sigma$ around the mean L$_{\rm\OIIIb}$-M$_{\rm B}$ relation and 
convolved this with the broad-band LF. We assumed a relation between L$_{\rm \OIIIb}$ 
and M$_{\rm B}$ derived for the SDSS type--1 sample by R08 and 
converted from a rest-frame wavelength of 2500 \AA\ to the B-band by adding 
0.2\footnote{Assuming the typical spectrum of type--1 AGN, 0.2 is the average color between 
the rest-frame wavelength of 2500\AA\ and 4400\AA\ (B-band).}, thus obtaining
\begin{equation}
\qquad \qquad \rm \log\left(\frac{L_{\rm \OIIIb}}{L_{\odot}}\right)= -0.38M_{\rm B}-0.42
\label{eq:MBLO3}
\end{equation}
with a scatter in L$_{\rm \OIIIb}$ at fixed continuum luminosity that is consistent with a 
log-normal scatter of width $\sigma$=0.36 dex.

By convolving the broad-band 
LF with the mean  L$_{\rm\OIIIb}$-M$_{\rm B}$ relation with its log-normal scatter (see Eq. (12) of R08),  
we obtain the \OIIIb\ LF model for type--1 AGN, which is shown as an orange line in 
the three upper panels of Fig. \ref{fig:fraction}, where additionally the SDSS type--1 AGN 
\OIIIb\ LF data points (derived directly from \OIIIb\ luminosities; R08) 
are shown as orange squares.

%\begin{equation}
%\Phi_{\rm conv}({\cal L}) = \int \Phi(M) \frac{1}{\sqrt{2\pi\sigma}}\exp\left[-\frac{({\cal L} - \overline{\cal L}(M))^2}{2\sigma^2}\right] d\,M,
%\Phi_{\rm conv}(L_{\rm [OIII]}) = \int \Phi(\Mi) \exp\left[-\frac{({\cal L} - \overline{\cal L}(\Mi))^2}{2\sigma^2}\right] d\Mi,
%\end{equation}
%where ${\cal L} = \log(L_{\rm [OIII]}/L_\odot)$ and $M = M_B$. 

By comparing this to the \OIIIb\ LF model derived in the previous section for type--2 AGN 
(black line in the same panels), we can directly estimate the type--2 AGN fraction 
in the three redshift bins. 
This is shown as a solid black line in the lower panels of Fig. \ref{fig:fraction}. 
%By using the LF we should be free from the selection bias due to 
%the absorption since the appropriate corrections have been already taken into account 
%in the computation of the LF itself.
We also computed the same quantity by considering the data points instead of the models.
In particular, for the type--2 AGN we used the LF data points, while for the type--1 AGN 
the only data available are from the SDSS (orange squares), and thus in the zCOSMOS 
regime we continued to use the extrapolated model.
The result of this second way of computing the type--2 AGN fraction, is shown with black circles 
(zCOSMOS) and blue diamonds (SDSS) in the lower panels of Fig. \ref{fig:fraction}.
As expected, the two methods are in good agreement and consistent within the errors, which 
are however, very large. 
%The only visible discrepancy is in the intermediate redshift bin, 
%where the data points are higher than one of the model (red line), as easily deducible looking at this 
%LF model fit compared to the data in the corresponding upper panel. 

For the 0.15$<z<$0.3 bin, we find that the type--2 fraction decreases with 
luminosity from $\sim$65\% (zCOSMOS data) 
to $\sim$50\% (SDSS data) going from L$_{\rm \OIIIb}$=10$^{6.2}$-10$^{8.2}$L$_{\odot}$ to brighter 
luminosities (L$_{\rm \OIIIb}$=10$^{8.2}$-10$^{9.2}$L$_{\odot}$). However, considering the errors and 
taking into account that the fractions derived from SDSS have to be considered as lower limits (see R08), 
the trend is also consistent with being constant with luminosity.
At higher redshift, the trend with luminosities is stronger, clearly showing a decreasing fraction of 
type--2 AGN with luminosity. 
%Moreover, these fractions are on average 
%higher than those in the first redshift bin \textbf{(e.g. at L$_{\rm \OIIIb} \sim 10^{7.5} L_{\odot}$ the fraction 
%increases from 73\% to 81\% going from z$\sim$ 0.2 to z$\sim$0.7)}, a hint for a possible increase 
%of the type--2 fraction with redshift, in agreement with e.g. \citet{Hasinger2008} and \citet{LaFranca2005}.
At 0.3$<z<$0.45 and 0.5$<z<$0.92, the type--2 fraction ranges from $\sim$80\% at 
L$_{\rm \OIIIb}$=10$^{6.2}$-10$^{8.5}$L$_{\odot}$ to $\sim$25\% in the SDSS regime 
(L$_{\rm \OIIIb}$=10$^{9.0}$-10$^{9.6}$L$_{\odot}$).
In contrast, we do not detect any clear trend with redshift. At fixed luminosity, e.g., 
L$_{\rm \OIIIb} \sim 10^{7.5}$ L$_{\odot}$, the fraction slight increases from 73\% to 81\% from z$\sim$ 0.2 to z$\sim$0.7. 
However, within the margins of error this is consistent with being constant.\\
We note that these results are based on two samples (zCOSMOS and SDSS) and 
that the different selection methods may affect the observed trend.
From the COSMOS data points alone, the type--2 fraction appears to be quite constant with luminosity 
in each redshift bin, although in a more limited luminosity range.

There has been a substantial amount of work to determine the 
obscured AGN fraction as a function of luminosity and redshift, leading sometimes to 
contradictory results. 
In the optical band, \citet{Simpson2005}, using a sample of type--1 AGN \OIIIb\ selected from 
the SDSS sample at 0.02$<z<$0.3, measured a decline in the type--2 fraction 
with luminosity. His fractions (represented by magenta squares in the first redshift bin) 
are broadly consistent with our (more uncertain) fractions.
In the X-ray band, where most of the work on this topic has been completed, 
several studies suggest that the fraction of obscured 
AGN decreases with luminosity and increases with z \citep{Ueda2003,Barger2005,LaFranca2005,
Treister2006a,Sazonov2007,Hasinger2008}, while 
other studies suggest that it is independent of both L and z \citep{Dwelly2006}, or that it is 
independent of z but not of L \citep{Ueda2003,Akylas2006,Gilli2007}. 

%The most relevant work done on this topic comes from X-ray studies, which suggest that 
%the fraction of obscured AGN is large at low X-ray luminosities and then decreases towards higher luminosities 
%\citep{Ueda2003,Barger2005,Treister2006a,Sazonov2007}. \\
The comparison of an optical type--2/(type--1+type--2) fraction with an X-ray obscured/(unobs+obs) 
fraction is always affected by limitations and differences related to the AGN classification in 
different bands. For example, Compton thick AGN may be missed by X-ray surveys \citep{LaMassa2009}, 
while the optical selection can fail to select  
AGN light diluted by their host galaxy.     
Moreover, the redshift range covered by X-ray samples usually extends to higher redshift  
than the optical ones. 
To overcome the difficulties in classifying either the 
optical or the X-ray bands, \citet{Hasinger2008} selected in the 2-10 keV band 
a sample of 1290 obscured AGN by combining both diagnostics. 
From this sample, he found a significant increase in the absorbed fraction with redshift and confirmed 
with higher quality statistics that there is a strong decline 
in the same fraction with X-ray luminosities. This decline can be described by an almost 
linear decrease from 80\% to 20\% in the luminosity range L$_X$=10$^{42}$-10$^{46}$.  
We show this trend with a dot-dashed green line in Fig. \ref{fig:fraction} after converting the X-ray 
to \OIIIb\ luminosity using the \citet{Netzer2006} relation. 
%In Fig. \ref{fig:fraction}, we also show the ratio of absorbed Compton-thick AGN to all Compton-thin AGN assumed 
%in the \citet{Gilli2007} X-ray background population synthesis model (red dotted line), which seems to 
%agree very well with our data at low luminosities, but overpredicts the fraction of type--2 AGN at high 
%luminosities.

\section{Summary and conclusions} \label{sec:conclusion}

We have presented the faintest optically selected sample of type--2 AGN to date. 
The sample, selected from the zCOSMOS survey, consists of 213 sources in the redshift 
range 0.15 $<z<$0.92, spanning the \OIIIb\ luminosity range 
$10^{5.5}$L$_{\odot}<$L$_{\rm \OIIIb}<$10$^{9.1}$ L$_{\odot}$. 
The only other sample at z$>$ 0.15 is one 
selected in a similar way from the SDSS by \citet{Zakamska2003}, which, 
however, covers significantly brighter luminosities ($10^{7.3}$L$_{\odot}<$L$_{\rm \OIIIb}<$10$^{10.1}$L$_{\odot}$).

Our sample has been selected using the first 10,000 spectra of the zCOSMOS dataset 
on the basis of their emission line properties.
In particular, we used the standard BPT diagrams (\OIIIb/\Hb\ versus \NII/\Ha\ and 
\OIIIb/\Hb\ versus \SII/\Ha) to isolate AGN in the redshift range 0.15$<z<$0.45 and the 
more recent diagnostic diagram \OIIIb/\Hb\ versus \OII/\Hb\ to extend the selection to 
higher redshift (0.5$<z<$0.92), after applying a cut to the S/N ratio of the involved 
lines.

Cross-checking the zCOSMOS emission-line sample with the XMM-COSMOS catalog, 
we found a significant incompleteness in the \blue\ diagnostic 
diagram used to select high redshift type--2 AGN (z$>$0.5). 
Our hypothesis is that LINERs
%, especially if they are affected by extinction on galactic scales, 
as well as composite AGN/SF sources can be misclassified by this diagram. 
%It is not surprising
%that the \blue\ diagram is sensitive to reddening effects given the use of the ratio between 
%two lines not close to each other (\OII\ and \Hb). 

For the selected type-2 AGN sample, we computed the \lOIIIb\ line luminosity function using 
the 1/V$_{\rm max}$ method.
The selection function takes into account (and corrects for) the sources that were not  
spectroscopically observed (but were in the photometric catalog) and those for 
which a secure spectroscopic identification has not been obtained. The correction is 
performed using a statistical weight associated with each galaxy that has secure redshift 
measurement. 
Since the sample was selected from a magnitude-limited sample, 
applying a criterion based on the line fluxes, the maximum volume V$_{\rm max}$ for 
each object has been estimated as the minimum between the volume $V_{max}(m_i)$ 
associated with the maximum apparent magnitude and 
the volumes $V_{max}(f_{i,l})$ associated with the minimum flux of the used lines.
We have extended the \lOIIIb\ LF to luminosities about 2 orders of magnitude fainter than those previously 
available (down to L$_{\rm \OIIIb}$=10$^{5.5} L_{\odot}$).  

%The comparison of our data with the model fit derived using the X-ray selected sample 
%from DC08 suggests that the LDDE model is a good representation of our data. \\
To enlarge the luminosity range, we combined our faint zCOSMOS sample with the
sample of bright type--2 AGN from the SDSS (R08) and  
found that the evolutionary model that represents the combined 
luminosity functions most accurately is an LDDE model. 
By fixing the evolutionary parameters (p1, p2, $\alpha$, z$_{\rm c,0}$, L$_{a}$) 
using the results obtained by DC08, we obtained
as best-fit model parameters $\gamma1$=0.56, $\gamma2$=2.42, and L$^*$=2.7$\times$10$^{41}$erg s$^{-1}$ with the normalization 
$\Phi^*_L$=1.08$\times$10$^{-5}$Mpc$^{-3}$.

Finally, by comparing the LF for type--2 and type--1 AGN (obtained by 
converting the broad-band LF from the VVDS into an \OIIIb\ LF), we constrained the type--2 quasar fraction as a function 
of luminosity. We found that the fraction of type--2 AGN is high at low \OIIIb\ luminosities 
and then decreases at higher luminosities, in agreement with that found from 
several studies in the X-ray band \citep{Ueda2003,Barger2005,Treister2006a,Sazonov2007,Gilli2007}.
The same decreasing trend with luminosity is found in all individual redshift bins, 
and we found on average a slightly higher fraction 
towards higher redshift 
(consistent however with being constant inside the errors).
In particular, we found that at 0.15$<z<$0.3 the fraction of type--2 AGN decreases with 
luminosity from $\sim$65\% 
to $\sim$50\%, 
while at 0.3$<z<$0.45, and 0.5$<z<$0.92 the type--2 fraction ranges from $\sim$80\% 
to $\sim$25\%. 
However, analysis of a sample uniformly selected across a wider luminosity range is needed to 
confirm these results, which are derived by combining two different samples (zCOSMOS and SDSS). 
We can not exclude that the observed trend could still be an artifact produced by the different 
selection functions in the two samples. From the COSMOS data points alone, 
the type--2 fraction seems to be quite constant with luminosity.

\begin{acknowledgements}
%AB wish to thank the anonymous referee for a detailed review of the manuscript that helped to improve the paper
AB wishes to thank the referee for the useful suggestions that significantly improved 
the manuscript.
AB also thank H. Netzer for the the stimulating discussion, R. Fassbender for helping to improve the manuscript and 
S. Savaglio for sharing the expertise.
This work is based on the zCOSMOS ESO Large Program Number 175.A-0839 and was  
supported by an INAF contract PRIN/2007/1.06.10.08, and an ASI grant ASI/COFIS/WP3110 I/026/07/0,
Partial support of this work is provided by NASA ADP NNX07AT02G, CONACyT 83564, and PAPIITIN110209.  
These results are also based in part on observations obtained with XMM-{\it Newton}, 
an ESA Science Mission with instruments and contributions directly funded by ESA Member 
States and the USA (NASA). 
%We thank the anonymous referee for his/her useful comments and suggestions. 
\end{acknowledgements}
%\bibliographystyle{aa}
%\bibliography{angi.bib}

\appendix
\section{Catalog Tables}
\label{tab}

%__________________________________________________ Table 
\begin{longtable}{lccrlccccc}
  \caption[]{\label{tab:catalog} zCOSMOS type--2 AGN.}   \\
  \hline\hline
  $ \rm Object \, ID $&$ \rm \hfill  \alpha_{J2000} \hfill $&$
  \hfill \rm  \delta_{J2000} \hfill $&$ \rm\hfill   z \hfill $&$ \rm flag^{(*)} $&$ \rm I_{AB} $&$
\rm log(L_{\rm [OIII]}) $&$ \rm Diag. Diagn. $&$ \rm Class $&$\rm XMM-ID^{(**)}$\\
  \hline
  \endfirsthead
  \caption{continued.}\\
  \hline\hline
  $ \rm Object \, ID $&$ \rm \hfill  \alpha_{J2000} \hfill $&$
  \hfill \rm  \delta_{J2000} \hfill $&$ \rm\hfill   z \hfill $&$ \rm flag^{(*)} $&$ \rm I_{AB} $&$
\rm log(L_{\rm [OIII]}) $&$ \rm Diag. Diagn. $&$ \rm Class $&$\rm XMM-ID^{(**)}$\\
   \hline
  \endhead
  \hline
  \endfoot
  $803032$ & $ 150.76768$ & $ 1.63525 $ & $0.622 $ & $23.5$ & $  22.19$ & $ 40.94$ &	    \blue\ &	   Sey2 & $ - $\\
$803488$ & $ 150.65017$ & $ 1.65705 $ & $0.587 $ & $ 4.5$ & $  20.43$ & $ 42.44$ &	    \blue\ &	   Sey2 & $ - $\\ 
$803690$ & $ 150.59690$ & $ 1.77130 $ & $0.305 $ & $ 4.5$ & $  22.44$ & $ 40.35$ &	    \redb\ &	   Sey2 & $ - $\\ 
$804921$ & $ 150.26019$ & $ 1.68284 $ & $0.368 $ & $ 3.5$ & $  19.75$ & $ 40.24$ &	    \reda\ &	   Sey2 & $ - $\\ 
$805007$ & $ 150.24253$ & $ 1.76639 $ & $0.623 $ & $ 4.5$ & $  20.63$ & $ 41.74$ &	    \blue\ &	   Sey2 & $ 104 $\\ 
$805085$ & $ 150.22834$ & $ 1.76728 $ & $0.350 $ & $ 3.5$ & $  20.09$ & $ 40.55$ &	    \reda\ &	   Sey2 & $ 385 $\\ 
$805561$ & $ 150.14424$ & $ 1.78938 $ & $0.530 $ & $ 3.5$ & $  20.50$ & $ 41.09$ &	    \blue\ &	   Sey2 & $ - $\\ 
$806212$ & $ 149.98870$ & $ 1.63793 $ & $0.776 $ & $ 4.5$ & $  22.35$ & $ 41.85$ &	    \blue\ &	   Sey2 & $ - $\\ 
$807990$ & $ 149.58416$ & $ 1.66021 $ & $0.569 $ & $ 4.5$ & $  22.44$ & $ 41.24$ &	    \blue\ &	   Sey2 & $ - $\\ 
$808045$ & $ 149.57083$ & $ 1.68372 $ & $0.549 $ & $24.5$ & $  21.84$ & $ 41.42$ &	    \blue\ &	   Sey2 & $ - $\\ 
$809516$ & $ 150.64840$ & $ 1.81752 $ & $0.357 $ & $ 4.1$ & $  22.13$ & $ 40.84$ &	    \reda\ &	   Sey2 & $ - $\\ 
$809527$ & $ 150.64657$ & $ 1.82997 $ & $0.414 $ & $ 4.1$ & $  21.93$ & $ 40.81$ &	    \reda\ &	   Sey2 & $ - $\\ 
$810119$ & $ 150.52413$ & $ 1.83500 $ & $0.268 $ & $ 9.5$ & $  22.41$ & $ 39.47$ &	    \reda\ &	   Sey2 & $ - $\\ 
$810775$ & $ 150.39021$ & $ 1.91442 $ & $0.844 $ & $ 4.5$ & $  21.58$ & $ 42.52$ &	    \blue\ &	   Sey2 & $ - $\\ 
$811278$ & $ 150.27067$ & $ 1.93005 $ & $0.834 $ & $ 3.5$ & $  22.42$ & $ 41.16$ &	    \blue\ &	   Sey2 & $ - $\\ 
$811623$ & $ 150.19950$ & $ 1.93154 $ & $0.168 $ & $ 3.5$ & $  21.31$ & $ 39.69$ &	    \redb\ &	   Sey2 & $ - $\\ 
$813133$ & $ 149.86204$ & $ 1.89481 $ & $0.444 $ & $ 4.5$ & $  18.95$ & $ 42.10$ &	    \reda\ &	   Sey2 & $ 293 $\\ 
$813250$ & $ 149.83058$ & $ 1.90213 $ & $0.730 $ & $ 4.5$ & $  21.62$ & $ 41.71$ &	    \blue\ &	   Sey2 & $ - $\\ 
$813693$ & $ 149.72622$ & $ 1.83715 $ & $0.674 $ & $ 2.1$ & $  21.38$ & $ 41.14$ &	    \blue\ &	   Sey2 & $ - $\\ 
$814251$ & $ 149.60093$ & $ 1.82646 $ & $0.433 $ & $ 4.5$ & $  21.92$ & $ 40.61$ &	    \reda\ &	   Sey2 & $ - $\\ 
$814284$ & $ 149.59289$ & $ 1.94331 $ & $0.250 $ & $ 3.5$ & $  22.14$ & $ 39.77$ &	    \redb\ &	   Sey2 & $ - $\\ 
$816490$ & $ 150.50326$ & $ 2.05877 $ & $0.368 $ & $ 3.5$ & $  19.58$ & $ 41.02$ &	    \reda\ &	   Sey2 & $ - $\\ 
$817444$ & $ 150.33577$ & $ 2.11693 $ & $0.537 $ & $ 4.5$ & $  20.13$ & $ 41.27$ &	    \blue\ &	   Sey2 & $ - $\\ 
$817672$ & $ 150.29147$ & $ 2.07307 $ & $0.426 $ & $ 3.5$ & $  19.97$ & $ 41.52$ &	    \reda\ &	   Sey2 & $ - $\\ 
$817907$ & $ 150.23947$ & $ 1.98371 $ & $0.511 $ & $ 4.1$ & $  20.53$ & $ 41.16$ &	    \blue\ &	   Sey2 & $60211$\\ 
$817943$ & $ 150.23060$ & $ 2.04859 $ & $0.830 $ & $ 4.5$ & $  22.45$ & $ 41.90$ &	    \blue\ &	   Sey2 & $ - $\\ 
$818788$ & $ 150.06034$ & $ 1.97416 $ & $0.336 $ & $ 3.5$ & $  22.47$ & $ 40.11$ &	    \reda\ &	   Sey2 & $ - $\\ 
$819003$ & $ 150.00781$ & $ 2.05307 $ & $0.257 $ & $ 4.5$ & $  22.50$ & $ 39.73$ &	    \redb\ &	   Sey2 & $ - $\\ 
$819431$ & $ 149.90112$ & $ 2.11145 $ & $0.773 $ & $ 3.5$ & $  20.91$ & $ 41.85$ &	    \blue\ &	   Sey2 & $ - $\\ 
$820316$ & $ 149.66920$ & $ 2.07405 $ & $0.340 $ & $ 4.5$ & $  19.06$ & $ 41.51$ &	    \reda\ &	   Sey2 & $417$\\ 
$820515$ & $ 149.62115$ & $ 2.02135 $ & $0.341 $ & $ 4.5$ & $  21.20$ & $ 40.61$ &	    \reda\ &	   Sey2 & $ - $\\ 
$820983$ & $ 149.52125$ & $ 2.07939 $ & $0.675 $ & $ 4.5$ & $  20.88$ & $ 41.83$ &	    \blue\ &	   Sey2 & $5307$\\ 
$823841$ & $ 150.31797$ & $ 2.23407 $ & $0.374 $ & $ 3.5$ & $  20.67$ & $ 40.34$ &	    \reda\ &	   Sey2 & $ 203 $\\ 
$824055$ & $ 150.26632$ & $ 2.16649 $ & $0.734 $ & $ 3.5$ & $  22.43$ & $ 41.12$ &	    \blue\ &	   Sey2 & $ - $\\ 
$824096$ & $ 150.25558$ & $ 2.16464 $ & $0.525 $ & $ 2.5$ & $  22.30$ & $ 40.77$ &	    \blue\ &	   Sey2 & $ - $\\ 
$824363$ & $ 150.20420$ & $ 2.27451 $ & $0.219 $ & $ 2.5$ & $  20.92$ & $ 39.33$ &	    \reda\ &	   Sey2 & $ - $\\ 
$824420$ & $ 150.19311$ & $ 2.21009 $ & $0.752 $ & $ 2.5$ & $  22.12$ & $ 41.04$ &	    \blue\ &	   Sey2 & $ - $\\ 
$824849$ & $ 150.09733$ & $ 2.17753 $ & $0.758 $ & $ 3.5$ & $  21.55$ & $ 41.33$ &	    \blue\ &	   Sey2 & $ - $\\ 
$825411$ & $ 149.99502$ & $ 2.23858 $ & $0.380 $ & $ 4.5$ & $  21.68$ & $ 39.82$ &	    \reda\ &	   Sey2 & $ - $\\ 
$826095$ & $ 149.85676$ & $ 2.27313 $ & $0.764 $ & $ 3.5$ & $  20.62$ & $ 41.76$ &	    \blue\ &	   Sey2 & $364$\\ 
$826475$ & $ 149.78190$ & $ 2.13905 $ & $0.355 $ & $ 4.5$ & $  18.92$ & $ 41.20$ &	    \reda\ &	   Sey2 & $63$\\ 
$827053$ & $ 149.66590$ & $ 2.29256 $ & $0.627 $ & $ 3.5$ & $  22.39$ & $ 40.82$ &	    \blue\ &	   Sey2 & $ - $\\ 
$830061$ & $ 150.42443$ & $ 2.33795 $ & $0.595 $ & $ 2.5$ & $  22.36$ & $ 40.41$ &	    \blue\ &	   Sey2 & $ - $\\ 
$831402$ & $ 150.16346$ & $ 2.43274 $ & $0.265 $ & $ 4.5$ & $  21.91$ & $ 40.10$ &	\reda\ \& \redb\ &	Sey2 & $ - $\\ 
$833656$ & $ 149.74156$ & $ 2.32745 $ & $0.790 $ & $ 3.5$ & $  22.45$ & $ 41.60$ &	    \blue\ &	   Sey2 & $ - $\\ 
$833683$ & $ 149.73616$ & $ 2.41912 $ & $0.570 $ & $ 3.5$ & $  22.36$ & $ 40.86$ &	    \blue\ &	   Sey2 & $ - $\\ 
$833705$ & $ 149.73153$ & $ 2.41389 $ & $0.668 $ & $ 3.5$ & $  20.97$ & $ 41.39$ &	    \blue\ &	   Sey2 & $ - $\\ 
$835655$ & $ 150.71046$ & $ 2.47737 $ & $0.360 $ & $ 3.5$ & $  18.79$ & $ 41.69$ &	    \reda\ &	   Sey2 & $ - $\\ 
$835863$ & $ 150.66372$ & $ 2.51845 $ & $0.403 $ & $24.5$ & $  21.40$ & $ 40.04$ &	    \reda\ &	   Sey2 & $ - $\\ 
$836868$ & $ 150.47703$ & $ 2.49409 $ & $0.679 $ & $ 4.5$ & $  20.78$ & $ 42.17$ &	    \blue\ &	   Sey2 & $ - $\\ 
$838362$ & $ 150.20448$ & $ 2.60782 $ & $0.410 $ & $ 3.1$ & $  22.07$ & $ 40.53$ &	    \reda\ &	   Sey2 & $ - $\\ 
$838560$ & $ 150.17137$ & $ 2.56402 $ & $0.502 $ & $ 3.5$ & $  21.02$ & $ 40.62$ &	    \blue\ &	   Sey2 & $121$\\ 
$841009$ & $ 149.71246$ & $ 2.57225 $ & $0.615 $ & $ 3.5$ & $  22.37$ & $ 41.04$ &	    \blue\ &	   Sey2 & $ - $\\ 
$841281$ & $ 149.65571$ & $ 2.60081 $ & $0.735 $ & $ 4.1$ & $  20.63$ & $ 42.71$ &	    \blue\ &	   Sey2 & $2076$\\ 
$841620$ & $ 149.57827$ & $ 2.49944 $ & $0.220 $ & $ 3.5$ & $  21.25$ & $ 39.71$ &	    \reda\ &	   Sey2 & $ - $\\ 
$844050$ & $ 150.49067$ & $ 2.63469 $ & $0.346 $ & $ 4.5$ & $  19.32$ & $ 40.84$ &	    \reda\ &	   Sey2 & $191$\\ 
$844969$ & $ 150.30461$ & $ 2.68244 $ & $0.741 $ & $ 4.5$ & $  22.28$ & $ 41.39$ &	    \blue\ &	   Sey2 & $ - $\\ 
$845167$ & $ 150.26257$ & $ 2.67084 $ & $0.270 $ & $ 3.5$ & $  19.43$ & $ 41.26$ &	\reda\ \& \redb\ &	Sey2 & $100$\\ 
$845190$ & $ 150.25882$ & $ 2.65369 $ & $0.267 $ & $ 2.5$ & $  21.87$ & $ 39.62$ &	    \reda\ &	   Sey2 & $ - $\\ 
$845769$ & $ 150.14062$ & $ 2.71089 $ & $0.777 $ & $ 3.5$ & $  22.44$ & $ 41.25$ &	    \blue\ &	   Sey2 & $ - $\\ 
$848207$ & $ 149.62741$ & $ 2.66164 $ & $0.358 $ & $ 3.5$ & $  19.83$ & $ 41.27$ &	    \reda\ &	   Sey2 & $ - $\\ 
$848413$ & $ 149.56764$ & $ 2.67188 $ & $0.221 $ & $ 4.5$ & $  18.65$ & $ 40.43$ &	\reda\ \& \redb\ &	Sey2 & $ - $\\ 
$848720$ & $ 149.49758$ & $ 2.65300 $ & $0.267 $ & $ 1.5$ & $  22.24$ & $ 39.37$ &	    \reda\ &	   Sey2 & $ - $\\ 
$800984$ & $ 150.28646$ & $ 1.62392 $ & $0.595 $ & $ 4.5$ & $  22.07$ & $ 41.94$ &	    \blue\ &  Sey2 cand & $ - $\\ 
$803226$ & $ 150.70539$ & $ 1.71696 $ & $0.570 $ & $ 4.5$ & $  22.40$ & $ 41.56$ &	    \blue\ &  Sey2 cand & $ - $\\ 
$803924$ & $ 150.51625$ & $ 1.68451 $ & $0.793 $ & $ 4.5$ & $  22.05$ & $ 41.78$ &	    \blue\ &  Sey2 cand & $ - $\\ 
$805300$ & $ 150.19023$ & $ 1.65969 $ & $0.516 $ & $ 4.5$ & $  21.41$ & $ 40.42$ &	    \blue\ &  Sey2 cand & $ - $\\ 
$807230$ & $ 149.76208$ & $ 1.76743 $ & $0.683 $ & $ 2.5$ & $  22.37$ & $ 41.00$ &	    \blue\ &  Sey2 cand & $ - $\\ 
$807785$ & $ 149.63644$ & $ 1.63659 $ & $0.667 $ & $ 9.5$ & $  22.32$ & $ 40.90$ &	    \blue\ &  Sey2 cand & $ - $\\ 
$811012$ & $ 150.33102$ & $ 1.87974 $ & $0.839 $ & $ 4.5$ & $  21.88$ & $ 42.26$ &	    \blue\ &  Sey2 cand & $ - $\\ 
$811024$ & $ 150.32867$ & $ 1.82685 $ & $0.811 $ & $ 4.5$ & $  22.23$ & $ 42.14$ &	    \blue\ &  Sey2 cand & $ - $\\ 
$811075$ & $ 150.31761$ & $ 1.92252 $ & $0.724 $ & $ 4.5$ & $  22.12$ & $ 41.50$ &	    \blue\ &  Sey2 cand & $ - $\\ 
$811519$ & $ 150.22358$ & $ 1.86679 $ & $0.672 $ & $ 3.5$ & $  22.12$ & $ 41.10$ &	    \blue\ &  Sey2 cand & $ - $\\ 
$811525$ & $ 150.22271$ & $ 1.80714 $ & $0.530 $ & $ 3.5$ & $  21.49$ & $ 40.71$ &	    \blue\ &  Sey2 cand & $2448$\\ 
$812173$ & $ 150.08485$ & $ 1.90615 $ & $0.838 $ & $ 4.1$ & $  22.46$ & $ 41.61$ &	    \blue\ &  Sey2 cand & $ - $\\ 
$813325$ & $ 149.81274$ & $ 1.82397 $ & $0.530 $ & $ 4.5$ & $  20.89$ & $ 41.08$ &	    \blue\ &  Sey2 cand & $5502$\\ 
$813609$ & $ 149.74226$ & $ 1.80049 $ & $0.693 $ & $ 4.5$ & $  22.48$ & $ 41.01$ &	    \blue\ &  Sey2 cand & $ - $\\ 
$813759$ & $ 149.71685$ & $ 1.86958 $ & $0.615 $ & $ 4.5$ & $  22.37$ & $ 40.65$ &	    \blue\ &  Sey2 cand & $ - $\\ 
$813833$ & $ 149.70129$ & $ 1.87056 $ & $0.787 $ & $ 3.5$ & $  21.85$ & $ 40.73$ &	    \blue\ &  Sey2 cand & $ - $\\ 
$817109$ & $ 150.40248$ & $ 2.07014 $ & $0.748 $ & $ 3.5$ & $  21.82$ & $ 41.08$ &	    \blue\ &  Sey2 cand & $ - $\\ 
$817802$ & $ 150.26492$ & $ 2.08250 $ & $0.888 $ & $ 4.5$ & $  22.25$ & $ 41.72$ &	    \blue\ &  Sey2 cand & $ - $\\ 
$818408$ & $ 150.13594$ & $ 2.12021 $ & $0.669 $ & $ 3.5$ & $  20.85$ & $ 41.23$ &	    \blue\ &  Sey2 cand & $ - $\\ 
$819142$ & $ 149.96681$ & $ 2.07602 $ & $0.715 $ & $ 4.5$ & $  22.36$ & $ 41.33$ &	    \blue\ &  Sey2 cand & $ - $\\ 
$820055$ & $ 149.73767$ & $ 2.06520 $ & $0.678 $ & $ 4.5$ & $  20.14$ & $ 41.56$ &	    \blue\ &  Sey2 cand & $5440$\\ 
$820087$ & $ 149.73082$ & $ 2.01584 $ & $0.647 $ & $ 4.5$ & $  21.75$ & $ 41.51$ &	    \blue\ &  Sey2 cand & $ - $\\ 
$820960$ & $ 149.52601$ & $ 2.02049 $ & $0.623 $ & $ 3.5$ & $  21.43$ & $ 41.08$ &	    \blue\ &  Sey2 cand & $5493$\\ 
$823458$ & $ 150.39592$ & $ 2.20760 $ & $0.672 $ & $ 3.5$ & $  20.83$ & $ 41.19$ &	    \blue\ &  Sey2 cand & $ - $\\ 
$823828$ & $ 150.31987$ & $ 2.25700 $ & $0.598 $ & $ 3.5$ & $  22.39$ & $ 41.04$ &	    \blue\ &  Sey2 cand & $ - $\\ 
$824005$ & $ 150.28031$ & $ 2.25339 $ & $0.665 $ & $ 3.5$ & $  21.90$ & $ 40.84$ &	    \blue\ &  Sey2 cand & $ - $\\ 
$824129$ & $ 150.24775$ & $ 2.18188 $ & $0.751 $ & $ 3.5$ & $  22.24$ & $ 41.17$ &	    \blue\ &  Sey2 cand & $ - $\\ 
$824914$ & $ 150.08787$ & $ 2.19456 $ & $0.688 $ & $ 3.5$ & $  20.80$ & $ 40.87$ &	    \blue\ &  Sey2 cand & $ - $\\ 
$825103$ & $ 150.04956$ & $ 2.24032 $ & $0.602 $ & $ 4.5$ & $  21.33$ & $ 41.98$ &	    \blue\ &  Sey2 cand & $60364$\\ 
$826547$ & $ 149.76593$ & $ 2.13280 $ & $0.830 $ & $ 4.5$ & $  21.53$ & $ 41.37$ &	    \blue\ &  Sey2 cand & $ - $\\ 
$829239$ & $ 150.58909$ & $ 2.31868 $ & $0.730 $ & $ 4.5$ & $  21.29$ & $ 41.02$ &	    \blue\ &  Sey2 cand & $ - $\\ 
$830321$ & $ 150.38336$ & $ 2.37199 $ & $0.851 $ & $ 4.5$ & $  22.32$ & $ 42.32$ &	    \blue\ &  Sey2 cand & $ - $\\ 
$830920$ & $ 150.26402$ & $ 2.38508 $ & $0.729 $ & $ 4.5$ & $  21.85$ & $ 41.71$ &	    \blue\ &  Sey2 cand & $ - $\\ 
$831156$ & $ 150.21556$ & $ 2.38531 $ & $0.837 $ & $ 3.5$ & $  22.45$ & $ 41.32$ &	    \blue\ &  Sey2 cand & $ - $\\ 
$831622$ & $ 150.11854$ & $ 2.41557 $ & $0.837 $ & $ 4.5$ & $  21.25$ & $ 42.42$ &	    \blue\ &  Sey2 cand & $ - $\\ 
$832586$ & $ 149.94662$ & $ 2.45866 $ & $0.782 $ & $ 9.3$ & $  22.25$ & $ 41.06$ &	    \blue\ &  Sey2 cand & $ - $\\ 
$833597$ & $ 149.75390$ & $ 2.29810 $ & $0.725 $ & $ 3.5$ & $  22.43$ & $ 40.81$ &	    \blue\ &  Sey2 cand & $ - $\\ 
$836763$ & $ 150.49719$ & $ 2.49495 $ & $0.672 $ & $ 4.5$ & $  22.43$ & $ 41.32$ &	    \blue\ &  Sey2 cand & $ - $\\ 
$837433$ & $ 150.37818$ & $ 2.53078 $ & $0.796 $ & $ 3.5$ & $  21.85$ & $ 40.97$ &	    \blue\ &  Sey2 cand & $ - $\\ 
$839230$ & $ 150.05230$ & $ 2.59478 $ & $0.696 $ & $ 4.5$ & $  22.37$ & $ 41.85$ &	    \blue\ &  Sey2 cand & $ - $\\ 
$839307$ & $ 150.03839$ & $ 2.56315 $ & $0.738 $ & $ 3.5$ & $  21.94$ & $ 40.86$ &	    \blue\ &  Sey2 cand & $ - $\\ 
$842052$ & $ 149.48318$ & $ 2.50662 $ & $0.702 $ & $ 4.5$ & $  22.37$ & $ 41.08$ &	    \blue\ &  Sey2 cand & $ - $\\ 
$843329$ & $ 150.63099$ & $ 2.66189 $ & $0.506 $ & $ 4.5$ & $  21.91$ & $ 42.44$ &	    \blue\ &  Sey2 cand & $ - $\\ 
$843389$ & $ 150.61933$ & $ 2.68042 $ & $0.510 $ & $ 3.5$ & $  21.70$ & $ 40.26$ &	    \blue\ &  Sey2 cand & $ - $\\ 
$844106$ & $ 150.48008$ & $ 2.69891 $ & $0.625 $ & $ 4.5$ & $  22.44$ & $ 40.95$ &	    \blue\ &  Sey2 cand & $ - $\\ 
$844618$ & $ 150.37585$ & $ 2.77923 $ & $0.500 $ & $ 4.5$ & $  22.15$ & $ 40.91$ &	    \blue\ &  Sey2 cand & $ - $\\ 
$846030$ & $ 150.09468$ & $ 2.74535 $ & $0.503 $ & $21.5$ & $  21.67$ & $ 40.45$ &	    \blue\ &  Sey2 cand & $ - $\\ 
$846516$ & $ 150.00840$ & $ 2.70455 $ & $0.581 $ & $ 2.1$ & $  20.88$ & $ 41.15$ &	    \blue\ &  Sey2 cand & $5572$\\ 
$847272$ & $ 149.85305$ & $ 2.74751 $ & $0.568 $ & $ 4.5$ & $  21.49$ & $ 41.29$ &	    \blue\ &  Sey2 cand & $ - $\\ 
$847277$ & $ 149.85273$ & $ 2.78911 $ & $0.598 $ & $ 4.5$ & $  22.31$ & $ 41.78$ &	    \blue\ &  Sey2 cand & $ - $\\ 
$848174$ & $ 149.63697$ & $ 2.67238 $ & $0.515 $ & $ 3.5$ & $  22.47$ & $ 40.29$ &	    \blue\ &  Sey2 cand & $ - $\\ 
$850262$ & $ 150.38865$ & $ 2.83712 $ & $0.816 $ & $ 3.5$ & $  22.35$ & $ 41.63$ &	    \blue\ &  Sey2 cand & $ - $\\ 
$851807$ & $ 149.87678$ & $ 2.81183 $ & $0.698 $ & $ 4.5$ & $  20.97$ & $ 40.93$ &	    \blue\ &  Sey2 cand & $ - $\\ 
$801297$ & $ 150.16557$ & $ 1.61078 $ & $0.170 $ & $ 4.5$ & $  21.78$ & $ 39.25$ &	    \redb\ &	  LINER & $ - $\\ 
$804277$ & $ 150.41975$ & $ 1.77573 $ & $0.361 $ & $ 3.5$ & $  18.91$ & $ 39.70$ &	    \reda\ &	  LINER & $ - $\\ 
$804573$ & $ 150.34526$ & $ 1.69938 $ & $0.309 $ & $ 3.5$ & $  19.15$ & $ 39.93$ &     \reda\ (both)  &   LINER & $ - $\\ 
$805283$ & $ 150.19297$ & $ 1.75240 $ & $0.266 $ & $ 3.5$ & $  17.98$ & $ 39.98$ &	    \reda\ &	  LINER & $ - $\\ 
$806922$ & $ 149.82336$ & $ 1.79156 $ & $0.433 $ & $ 4.5$ & $  21.79$ & $ 40.42$ &	    \reda\ &	  LINER & $ - $\\ 
$807417$ & $ 149.72088$ & $ 1.69172 $ & $0.387 $ & $24.5$ & $  21.48$ & $ 40.59$ &	\redb\ (both) &   LINER & $ - $\\ 
$807941$ & $ 149.59828$ & $ 1.76895 $ & $0.356 $ & $ 3.5$ & $  22.11$ & $ 40.29$ &	    \reda\ &	  LINER & $ - $\\ 
$808018$ & $ 149.57618$ & $ 1.64344 $ & $0.284 $ & $ 2.5$ & $  18.75$ & $ 40.22$ &	    \reda\ &	  LINER & $ - $\\ 
$808251$ & $ 149.52226$ & $ 1.67913 $ & $0.324 $ & $ 4.5$ & $  22.26$ & $ 39.69$ &	    \reda\ &	  LINER & $ - $\\ 
$810929$ & $ 150.35191$ & $ 1.94829 $ & $0.350 $ & $ 3.5$ & $  22.44$ & $ 39.87$ &	\reda\ (both) &   LINER & $ - $\\ 
$810944$ & $ 150.34825$ & $ 1.94898 $ & $0.347 $ & $ 4.5$ & $  18.66$ & $ 39.88$ &	    \reda\ &	  LINER & $ - $\\ 
$811036$ & $ 150.32690$ & $ 1.95576 $ & $0.345 $ & $ 3.5$ & $  21.99$ & $ 39.83$ &	    \reda\ &	  LINER & $ - $\\ 
$811115$ & $ 150.30746$ & $ 1.91255 $ & $0.248 $ & $ 3.5$ & $  20.53$ & $ 40.13$ &	    \redb\ &	  LINER & $ - $\\ 
$812330$ & $ 150.04893$ & $ 1.94878 $ & $0.439 $ & $ 1.5$ & $  21.78$ & $ 39.81$ &	    \reda\ &	  LINER & $ - $\\ 
$812538$ & $ 150.00230$ & $ 1.86215 $ & $0.281 $ & $ 4.5$ & $  20.85$ & $ 40.69$ &	    \reda\ &	  LINER & $ - $\\ 
$812596$ & $ 149.98805$ & $ 1.82299 $ & $0.342 $ & $ 3.5$ & $  19.16$ & $ 39.98$ &	    \reda\ &	  LINER & $ - $\\ 
$812632$ & $ 149.97776$ & $ 1.83432 $ & $0.371 $ & $ 3.5$ & $  20.95$ & $ 40.24$ &	    \redb\ &	  LINER & $ - $\\ 
$812882$ & $ 149.91731$ & $ 1.86815 $ & $0.446 $ & $ 4.5$ & $  21.71$ & $ 40.84$ &	    \reda\ &	  LINER & $ - $\\ 
$813199$ & $ 149.84753$ & $ 1.92329 $ & $0.372 $ & $ 4.5$ & $  19.58$ & $ 39.97$ &	    \reda\ &	  LINER & $ - $\\ 
$813816$ & $ 149.70356$ & $ 1.80179 $ & $0.202 $ & $ 4.5$ & $  21.89$ & $ 39.81$ &	    \redb\ &	  LINER & $ - $\\ 
$813888$ & $ 149.68669$ & $ 1.85241 $ & $0.341 $ & $ 3.5$ & $  22.24$ & $ 40.02$ &	    \reda\ &	  LINER & $ - $\\ 
$814131$ & $ 149.62829$ & $ 1.82954 $ & $0.398 $ & $ 2.5$ & $  21.56$ & $ 39.54$ &	    \reda\ &	  LINER & $ - $\\ 
$816988$ & $ 150.41995$ & $ 1.97551 $ & $0.310 $ & $ 3.5$ & $  20.24$ & $ 39.49$ &	    \reda\ &	  LINER & $ - $\\ 
$816998$ & $ 150.41833$ & $ 2.08515 $ & $0.425 $ & $ 4.5$ & $  19.99$ & $ 41.24$ &	    \reda\ &	  LINER & $2195$\\ 
$818160$ & $ 150.18249$ & $ 2.03933 $ & $0.347 $ & $ 3.5$ & $  20.27$ & $ 39.59$ &	    \reda\ &	  LINER & $ - $\\ 
$818225$ & $ 150.17228$ & $ 2.00605 $ & $0.310 $ & $ 3.5$ & $  19.00$ & $ 39.62$ &	    \reda\ &	  LINER & $ - $\\ 
$818312$ & $ 150.15145$ & $ 1.96550 $ & $0.361 $ & $ 2.5$ & $  20.27$ & $ 40.00$ &	    \reda\ &	  LINER & $ - $\\ 
$818453$ & $ 150.12748$ & $ 2.11223 $ & $0.360 $ & $ 2.5$ & $  22.38$ & $ 39.57$ &	    \reda\ &	  LINER & $ - $\\ 
$818456$ & $ 150.12730$ & $ 2.08187 $ & $0.382 $ & $ 4.5$ & $  21.93$ & $ 40.01$ &	    \reda\ &	  LINER & $ - $\\ 
$818518$ & $ 150.11550$ & $ 1.98236 $ & $0.361 $ & $ 2.5$ & $  20.45$ & $ 39.96$ &	    \reda\ &	  LINER & $ - $\\ 
$818607$ & $ 150.10105$ & $ 1.99492 $ & $0.373 $ & $ 2.5$ & $  21.22$ & $ 39.80$ &	    \reda\ &	  LINER & $ - $\\ 
$818868$ & $ 150.04240$ & $ 2.12580 $ & $0.340 $ & $ 4.5$ & $  21.08$ & $ 40.29$ &	    \reda\ &	  LINER & $ - $\\ 
$818949$ & $ 150.02084$ & $ 2.05949 $ & $0.373 $ & $ 3.5$ & $  19.40$ & $ 40.07$ &	    \reda\ &	  LINER & $70220$\\ 
$819090$ & $ 149.98082$ & $ 2.05774 $ & $0.358 $ & $ 4.5$ & $  21.74$ & $ 40.38$ &	    \reda\ &	  LINER & $ - $\\ 
$819241$ & $ 149.94342$ & $ 2.09623 $ & $0.356 $ & $ 3.5$ & $  21.47$ & $ 40.11$ &	    \reda\ &	  LINER & $ - $\\ 
$819294$ & $ 149.93138$ & $ 2.04602 $ & $0.442 $ & $ 3.5$ & $  21.91$ & $ 40.03$ &	    \reda\ &	  LINER & $ - $\\ 
$819347$ & $ 149.92088$ & $ 2.03123 $ & $0.356 $ & $ 4.5$ & $  18.98$ & $ 40.54$ &	    \reda\ &	  LINER & $ - $\\ 
$819596$ & $ 149.86483$ & $ 2.00328 $ & $0.417 $ & $ 3.5$ & $  22.30$ & $ 40.30$ &	    \redb\ &	  LINER & $ - $\\ 
$819667$ & $ 149.84478$ & $ 2.01555 $ & $0.253 $ & $ 3.5$ & $  20.47$ & $ 40.02$ &	\reda\ \& \redb\ &     LINER & $ - $\\ 
$819739$ & $ 149.82717$ & $ 2.03881 $ & $0.361 $ & $ 4.5$ & $  22.15$ & $ 40.35$ &	    \reda\ &	  LINER & $ - $\\ 
$819813$ & $ 149.80448$ & $ 2.09796 $ & $0.283 $ & $ 2.5$ & $  22.15$ & $ 39.80$ &	    \reda\ &	  LINER & $ - $\\ 
$820454$ & $ 149.63649$ & $ 2.02012 $ & $0.354 $ & $ 4.5$ & $  19.09$ & $ 40.04$ &	    \reda\ &	  LINER & $ - $\\ 
$820664$ & $ 149.58821$ & $ 2.08737 $ & $0.285 $ & $ 2.5$ & $  20.45$ & $ 39.65$ &	    \reda\ &	  LINER & $ - $\\ 
$823616$ & $ 150.36183$ & $ 2.26476 $ & $0.255 $ & $ 2.5$ & $  18.81$ & $ 39.29$ &	    \reda\ &	  LINER & $ - $\\ 
$824083$ & $ 150.25961$ & $ 2.20985 $ & $0.372 $ & $ 2.5$ & $  20.16$ & $ 39.93$ &	    \reda\ &	  LINER & $ - $\\ 
$824405$ & $ 150.19694$ & $ 2.28779 $ & $0.441 $ & $ 3.5$ & $  20.70$ & $ 40.01$ &	    \reda\ &	  LINER & $ - $\\ 
$824856$ & $ 150.09587$ & $ 2.25361 $ & $0.321 $ & $ 4.5$ & $  21.88$ & $ 40.35$ &	    \redb\ &	  LINER & $ - $\\ 
$825006$ & $ 150.06764$ & $ 2.24299 $ & $0.345 $ & $ 4.5$ & $  18.97$ & $ 40.31$ &	\reda\ \& \redb\ &     LINER & $ - $\\ 
$825708$ & $ 149.93176$ & $ 2.28066 $ & $0.413 $ & $ 3.5$ & $  22.30$ & $ 39.83$ &	    \redb\ &	  LINER & $ - $\\ 
$825904$ & $ 149.89498$ & $ 2.20841 $ & $0.344 $ & $ 4.5$ & $  19.96$ & $ 40.37$ &	\reda\ \& \redb\ &     LINER & $ - $\\ 
$826135$ & $ 149.84931$ & $ 2.13400 $ & $0.373 $ & $ 2.5$ & $  22.08$ & $ 39.45$ &	    \redb\ &	  LINER & $ - $\\ 
$826201$ & $ 149.83799$ & $ 2.20825 $ & $0.384 $ & $ 3.5$ & $  22.13$ & $ 39.92$ &	    \redb\ &	  LINER & $ - $\\ 
$826888$ & $ 149.69905$ & $ 2.24733 $ & $0.220 $ & $ 3.5$ & $  21.74$ & $ 39.91$ &	    \reda\ &	  LINER & $ - $\\ 
$827400$ & $ 149.59808$ & $ 2.15839 $ & $0.379 $ & $ 2.5$ & $  21.41$ & $ 39.34$ &	    \redb\ &	  LINER & $ - $\\ 
$827762$ & $ 149.50944$ & $ 2.23188 $ & $0.282 $ & $ 3.5$ & $  18.27$ & $ 40.30$ &	    \reda\ &	  LINER & $ - $\\ 
$827818$ & $ 149.49485$ & $ 2.28065 $ & $0.305 $ & $ 4.5$ & $  18.80$ & $ 40.36$ &	    \reda\ &	  LINER & $ - $\\ 
$830098$ & $ 150.41909$ & $ 2.33659 $ & $0.430 $ & $ 4.5$ & $  21.57$ & $ 40.67$ &	    \reda\ &	  LINER & $ - $\\ 
$830317$ & $ 150.38441$ & $ 2.39127 $ & $0.374 $ & $ 4.5$ & $  19.79$ & $ 40.67$ &	    \reda\ &	  LINER & $ - $\\ 
$830802$ & $ 150.28584$ & $ 2.35820 $ & $0.340 $ & $ 4.5$ & $  21.06$ & $ 40.37$ &	    \redb\ &	  LINER & $ - $\\ 
$831857$ & $ 150.08065$ & $ 2.39024 $ & $0.353 $ & $ 3.5$ & $  18.76$ & $ 40.14$ &	    \reda\ &	  LINER & $ - $\\ 
$832130$ & $ 150.03523$ & $ 2.41512 $ & $0.371 $ & $ 2.5$ & $  20.58$ & $ 39.76$ &	    \reda\ &	  LINER & $ - $\\ 
$832140$ & $ 150.03337$ & $ 2.38071 $ & $0.342 $ & $ 1.5$ & $  22.36$ & $ 39.53$ &	    \reda\ &	  LINER & $ - $\\ 
$832153$ & $ 150.03082$ & $ 2.31296 $ & $0.437 $ & $ 3.5$ & $  21.23$ & $ 40.39$ &	    \reda\ &	  LINER & $ - $\\ 
$832425$ & $ 149.98074$ & $ 2.34280 $ & $0.282 $ & $ 1.5$ & $  22.01$ & $ 39.58$ &	    \reda\ &	  LINER & $ - $\\ 
$832819$ & $ 149.90313$ & $ 2.44337 $ & $0.342 $ & $ 2.5$ & $  21.52$ & $ 39.88$ &	    \reda\ &	  LINER & $ - $\\ 
$832902$ & $ 149.88401$ & $ 2.45838 $ & $0.333 $ & $ 3.5$ & $  19.41$ & $ 40.57$ &	\reda\ \& \redb\ &     LINER & $ - $\\ 
$832913$ & $ 149.88279$ & $ 2.34362 $ & $0.361 $ & $ 4.5$ & $  20.16$ & $ 39.65$ &	    \reda\ &	  LINER & $ - $\\ 
$833279$ & $ 149.81184$ & $ 2.44122 $ & $0.426 $ & $24.5$ & $  21.05$ & $ 40.07$ &	    \reda\ &	  LINER & $ - $\\ 
$833562$ & $ 149.76103$ & $ 2.38529 $ & $0.354 $ & $ 3.1$ & $  21.80$ & $ 40.27$ &	    \reda\ &	  LINER & $ - $\\ 
$833599$ & $ 149.75357$ & $ 2.39981 $ & $0.347 $ & $ 2.1$ & $  21.81$ & $ 39.63$ &	    \reda\ &	  LINER & $ - $\\ 
$833627$ & $ 149.74729$ & $ 2.34573 $ & $0.373 $ & $ 4.5$ & $  18.32$ & $ 40.27$ &	    \reda\ &	  LINER & $ - $\\ 
$834427$ & $ 149.59079$ & $ 2.42824 $ & $0.445 $ & $ 3.5$ & $  21.15$ & $ 40.54$ &	    \reda\ &	  LINER & $ - $\\ 
$836171$ & $ 150.60392$ & $ 2.54711 $ & $0.372 $ & $ 3.5$ & $  20.29$ & $ 39.73$ &	    \reda\ &	  LINER & $ - $\\ 
$837364$ & $ 150.39102$ & $ 2.50478 $ & $0.220 $ & $ 2.5$ & $  22.12$ & $ 39.30$ &	    \reda\ &	  LINER & $ - $\\ 
$837373$ & $ 150.38966$ & $ 2.59289 $ & $0.361 $ & $ 3.5$ & $  21.12$ & $ 40.32$ &	    \reda\ &	  LINER & $ - $\\ 
$837378$ & $ 150.38847$ & $ 2.57608 $ & $0.361 $ & $ 3.5$ & $  21.95$ & $ 40.12$ &	    \reda\ &	  LINER & $ - $\\ 
$837475$ & $ 150.37069$ & $ 2.52179 $ & $0.348 $ & $ 1.5$ & $  21.35$ & $ 39.49$ &	    \reda\ &	  LINER & $ - $\\ 
$837485$ & $ 150.36911$ & $ 2.61409 $ & $0.434 $ & $ 2.5$ & $  21.76$ & $ 39.89$ &	    \reda\ &	  LINER & $ - $\\ 
$837768$ & $ 150.31518$ & $ 2.51697 $ & $0.361 $ & $ 2.5$ & $  21.31$ & $ 39.69$ &	    \redb\ &	  LINER & $ - $\\ 
$838271$ & $ 150.22023$ & $ 2.52454 $ & $0.376 $ & $ 2.5$ & $  20.52$ & $ 39.52$ &	    \reda\ &	  LINER & $ - $\\ 
$838310$ & $ 150.21530$ & $ 2.60187 $ & $0.409 $ & $ 1.5$ & $  19.27$ & $ 40.06$ &	\reda\ \& \redb\ &     LINER & $ - $\\ 
$838415$ & $ 150.19491$ & $ 2.49020 $ & $0.376 $ & $ 3.5$ & $  21.09$ & $ 39.52$ &	\reda\ \& \redb\ &     LINER & $ - $\\ 
$839265$ & $ 150.04681$ & $ 2.48205 $ & $0.307 $ & $ 3.5$ & $  21.67$ & $ 39.84$ &	    \redb\ &	  LINER & $ - $\\ 
$839410$ & $ 150.02198$ & $ 2.56129 $ & $0.438 $ & $ 2.5$ & $  22.05$ & $ 40.00$ &	    \reda\ &	  LINER & $ - $\\ 
$839422$ & $ 150.01910$ & $ 2.59541 $ & $0.307 $ & $ 3.5$ & $  19.99$ & $ 39.70$ &	    \reda\ &	  LINER & $ - $\\ 
$839646$ & $ 149.97569$ & $ 2.46143 $ & $0.346 $ & $ 4.5$ & $  18.81$ & $ 40.57$ &	    \reda\ &	  LINER & $60305$\\ 
$840078$ & $ 149.91214$ & $ 2.56036 $ & $0.413 $ & $ 2.5$ & $  20.15$ & $ 39.94$ &	\reda\ \& \redb\ &     LINER & $ - $\\ 
$841692$ & $ 149.56105$ & $ 2.56277 $ & $0.247 $ & $ 1.5$ & $  22.36$ & $ 39.22$ &	    \redb\ &	  LINER & $ - $\\ 
$841723$ & $ 149.55378$ & $ 2.46013 $ & $0.375 $ & $ 3.5$ & $  22.02$ & $ 40.43$ &	\reda\ \& \redb\ &     LINER & $ - $\\ 
$842079$ & $ 149.47718$ & $ 2.58237 $ & $0.419 $ & $ 4.5$ & $  21.03$ & $ 40.65$ &	\reda\ \& \redb\ &     LINER & $2578$\\ 
$844873$ & $ 150.32940$ & $ 2.67168 $ & $0.430 $ & $ 4.5$ & $  21.72$ & $ 40.68$ &	    \redb\ &	  LINER & $ - $\\ 
$845172$ & $ 150.26220$ & $ 2.65797 $ & $0.348 $ & $ 4.5$ & $  21.60$ & $ 40.27$ &	    \redb\ &	  LINER & $ - $\\ 
$845649$ & $ 150.16024$ & $ 2.69930 $ & $0.305 $ & $ 3.5$ & $  19.00$ & $ 40.14$ &	    \reda\ &	  LINER & $ - $\\ 
$845945$ & $ 150.10801$ & $ 2.69555 $ & $0.350 $ & $ 2.5$ & $  19.37$ & $ 40.05$ &	    \reda\ &	  LINER & $ - $\\ 
$846070$ & $ 150.08940$ & $ 2.70716 $ & $0.355 $ & $ 3.5$ & $  22.02$ & $ 39.73$ &	\reda\ \& \redb\ &     LINER & $ - $\\ 
$846146$ & $ 150.07553$ & $ 2.71314 $ & $0.432 $ & $ 2.5$ & $  21.26$ & $ 39.71$ &	    \reda\ &	  LINER & $ - $\\ 
$848329$ & $ 149.58770$ & $ 2.76141 $ & $0.353 $ & $ 3.5$ & $  18.83$ & $ 39.95$ &	    \reda\ &	  LINER & $ - $\\ 

\end{longtable}
\begin{description}
  \item[$^{(*)}$] confidence level for the assigned redshift considering also the consistency with 
  the photometric redshift. It ranges from 1 (low quality) to 4.5 (high quality); a 2 is added in front when 
  the object has been observed as secondary target, for more details see \citet{Lilly2007,Lilly2009}. 
  \item[$^{(**)}$] see \citet{Cappelluti2009}.
 \end{description}
%--------------------------------------------------

\end{document}